\documentclass{revtex4-1}
\usepackage{graphicx}
\usepackage{mathtools}
\usepackage{natbib}
\usepackage{physics}
\usepackage{url}
\usepackage{color}
\usepackage{amsmath}
\usepackage{amssymb}
\usepackage{abbreviations} 
\usepackage{caption}

\newcommand{\ind}[1]{\textrm{#1}}
\newcommand{\indt}[1]{\textrm{\tiny#1}}
\newcommand{\oder}[2]{\frac{d #1}{d #2}}

\newcommand{\beq}{\begin{equation}}
\newcommand{\eeq}{\end{equation}}
\newcommand{\fracp}[2]{\left(\frac{#1}{#2}\right)}

\begin{document}
\setcounter{page}{0}
\title{The reconfinement of AGN jets}
\author{ J.T. \surname{Bamford} and S.S. \surname{Komissarov}}
\email{s.s.komissarov@leeds.ac.uk}
\affiliation{Department of Applied Mathematics\\
The University of Leeds\\ Leeds, LS29JT, UK}

\begin{abstract}
In this paper we study the reconfinement of initially freely-expanding unmagnetised relativistic jets by the pressure of non-uniform external gas using numerical approach.  The results are compared with the simple semi-analytic model proposed by Komissarov \& Falle \cite{KF-97}. In particular, we explore the reconfinement in power-law atmospheres and in the King atmosphere, which describes the X-ray coronas of giant elliptic galaxies.  The results show significant deviations from the KF model, which systematically underestimates the reconfinement scale.   For the power-law atmospheres $p_\ind{ext}\propto z^{-\kappa}$, the disagreement  increases with  $\kappa$, exceeding two orders of magnitude for $\kappa=1.5$. For the King model, strong deviations are found on the outskirts of the atmosphere, where the distribution approaches a power law.  However for jets reconfined inside the core, the reconfinement scale is increased only by the factor of two. When the King model is modified by adding a central cusp, this has little impact on the jets which are reconfined outside of the cusp region but inside the cusp the reconfinement scale significantly reduces.      
\end{abstract}

\maketitle

\section{Introduction}
\label{sec:intro}

One of the key features of active galactic nuclei (AGN) is the generation of highly collimated outflows. These jets are often bright sources of electromagnetic radiation in various energy bands, from radio to gamma-rays. Their interaction with the surrounding gas normally creates huge bubbles of radio-emitting plasma, known as radio lobes, and sends strong shock waves through the gas. 
These may greatly influence the dynamics of this gas and related processes such as the galactic star formation \citep{BBR-84,BHK12}.   

Remarkably, the AGN jets can extend well beyond their parent galaxies and reach distances up to one Mpc ($3\times10^{24}$cm) from the AGN.  Inside the AGN their initial radius may well be as small as the gravitational radius of the central black hole, which is only  $\sim 3\times 10^{14}\,$cm for a $10^9 M_\odot$ black hole \cite{Bz}. Thus the jets may cross distances up to a few billion initial radii and remain unscathed.  To put things into perspective, for an aircraft jet engine this would correspond to a jet extending all the way to the Moon. In reality, the terrestrial jets  get destroyed by dynamical instabilities over much shorter length scales, not exceeding even a hundred initial radii.   

This seemingly outstanding stability of AGN (and other cosmic) jets may have a rather simple explanation based on the well-known fact that these jets are not perfectly collimated and presumably highly supersonic. In fact, their radius increases with distance from the AGN by up to few million times. Provided the jet opening angle $\theta_\ind{j}$ exceeds the Mach angle, $\theta_\ind{m}$, no perturbation can involve the whole of the jet, which becomes causally-disconnected and hence globally stable.   Such conditions are naturally produced in freely expanding jets, whose asymptotic opening angle is determined by the initial Mach number but the downstream Mach number grows unlimited as the jet plasma cools via the expansion.  This argument is easily extendable to the case of magnetised jets \cite{MSO-08,PK-15}.    

Once a free jet has reached its asymptotic speed and opening angle, its pressure decreases with the distance from the origin as $p_\ind{j} \propto z^{-2\gamma}$, where $\gamma$ is the ratio of specific heats. This suggests that jets become asymptotically freely expanding when the external pressure decreases faster, e.g. $p_\ind{ext} \propto z^{-\kappa}$ with $\kappa >2\gamma$ for a power-law atmosphere \cite{sanders-83}. For $\kappa < 2\gamma$ the external pressure can never be ignored completely.  Moreover, the outcome depends on whether $\kappa<2$ or $2<\kappa<2\gamma$. 

For $\kappa<2$ one may expect the jet pressure to stay matched with that of the external gas. Indeed, for such pressure-matched jets $\theta_\ind{m}/\theta_\ind{j}\propto z^{(2-\kappa)/2}$ increases with the distance from the origin and hence the jet become increasingly causally-connected \cite{PK-15}. Although this implies the self-consistency of the steady-state model, such jets are vulnerable to instabilities. In particularly, the simulations by \cite{PK-15} show that the growth rates of magnetic kink modes decrease with $\kappa$ and vanish in the limit $\kappa\to 2$, with the exception of the central core which remains self-collimated and unstable.  

For $2<\kappa<2\gamma$,  the model of pressure-matched jet predicts that $\theta_\ind{m}/\theta_\ind{j}$ decreases with the distance and hence the jet cannot actually maintain the pressure-matching via sound waves.     
In this case the external pressure will drive a shock wave into the expanding jet. As the ram pressure of freely-expanding jets, $p_\ind{j,ram}\propto z^{-2}$, decreases slower than the external pressure one would expect the recollimation shock never to reach the jet axis but to split it into a freely-expanding core and a shocked outer shell instead.  Based on the causality argument, one would expect such jets to be globally stable. 

Using various observational estimates of the external pressure of AGN jets, \cite{phinney-83} concluded that it can be approximated by a power law with $\kappa\approx 2$.  However, it is unlikely that a single power law describes the external pressure distribution from $z=10^{14}$cm to $z=10^{24}$cm, as we expect very different physical processes to shape the external pressure on such different scales.  In fact, we know that the pressure distribution of galactic X-ray coronal gas is much flatter \citep[e.g.][]{Mathews}.  Outside of the galaxy, it may flatten furthermore if the galaxy is a member of a rich cluster of galaxies.  

On the contrary, near the AGN where the gravity of the central supermassive black hole dominates, one would expect a steeper profile.  For a polytropic atmosphere of a central mass, one has $\kappa=\gamma/(\gamma-1)$,
which is higher than 2 when $1<\gamma<2$. For a spherical adiabatic wind, $\kappa=2\gamma$, which is also 
steeper than the critical one.  For the Bondi accretion $\kappa =3\gamma/2$, which is still larger than 2 for 
$\gamma>4/3$.  For such a steep profile of the external pressure one may expect AGN jets to become freely-expanding with a thin shocked boundary layer.   When such a jet enters the flatter outer sections of the external gas,
e.g. corresponding to the galactic coronal gas, the recollimation shock may reach the jet axis and hence become a reconfinement shock.  

The plasma compression and dissipation of the bulk motion energy at shocks imply enhanced emissivity and shocks have been connected with bright knots of AGN and other astrophysical jets. In particular, the so-called ``superluminal'' knots of AGN jets, often observed on the pc-scale using VLBI radio telescopes,  are usually associated with non-stationary shocks induced by the variability of the central engine and travel down these jets. 
On the kpc-scale, the proper motion of the jet knots is usually unknown and hence it is hard to tell if they are also moving or stationary features instead.  However, they often display quasi-periodic structure reminiscent of the chain of secondary shocks associated with the reconfinement process \citep[e.g.][]{BHL-94,WF-85,FW-85}.  An additional impulse to the theoretical studies of the recollimation shocks was given by the  realisation that in addition to the superluminal knots, VLBI images of AGN jets also include quasi-stationary features. The recent systematic observations of large samples of VLBI-jets show that such features are quite common  \citep[e.g.][]{JMM-17}.  
The issues of location, geometry, dissipation efficiency and emission from the recollimation and reconfinement shocks of steady-state jets have been the subject of numerous theoretical studies  \citep[e.g.][]{wilson-87,DM-88,DP-93,KF-97,SAK-06,NS-09,Nalewajko-12,BL-07,BL-09,KB-12,KB-12a,KB-15}, with applications both to AGN and GRB jets.

\cite{PK-15} pointed out another important aspect of the reconfinement process -- the transition from the stable propagation regime to the unstable one.  Indeed, once the jets get reconfined by the external pressure and hence become causally connected once again, they may develop either Kelvin-Helmholtz (KHI) or/and hydromagnetic instabilities.  \cite{MM13} and \cite{MAP17} also argued that the reconfinement process itself can be accompanied by the Rayleigh-Taylor instability (RTI) driven by the centrifugal acceleration associated with the curved streamlines of the flow.  They concluded that for the instability to develop the jet must be heavier compared to its immediate surrounding, which is expected to be the case when jets are reconfined by the pressure of their own cocoons \cite[cf.][]{TK-17}.   However, \cite{GK-18NatAs} carried out 3D time-dependent simulations of relativistic jets undergoing reconfinement by the external gas pressure and observed a rapid loss of stability and quick transition to fully turbulent regime both in the case of heavy and light jets.  They interpreted the instability not as the inertial RTI but as the more general centrifugal instability (CFI). \cite{GK-18} derived the relativistic extension of the Rayleigh instability criterion for CFI and verified it in the simple case of rotating fluid.

In the context of AGN jets, early reconfinement and transition to the turbulent regime is consistent with the observed morphology of type-I sources in the Fanaroff-Riley classification scheme  \cite{FR-74,PK-15,GK-18NatAs}. In this regard, it is important to be able to predict the location of the reconfinement point (RP), the point where the reconfinement shock reaches the jet axis.  \cite{F-91} derived a simple differential equation for the geometry of the reconfinement shock in an initially free-expanding cold non-relativistic jet. The main simplifying assumption of the derivation is that the pressure across the shocked outer layer of the jet is constant and equal to that of the external gas --  this is similar to the approximation made by Kompaneets in modelling strong explosions \cite{Komp-60}. This approach was extended by Komissarov \& Falle \citep[][KF]{KF-97} to include relativistic flows. 

In the case of power-law atmospheres, the shock equation allows a simple analytic solution and hence gives the position of the reconfinement point as a relatively simple function of the jet power and few other parameters.  However the reconfinement shock is highly oblique and hence the flow of the shocked outer layer remains supersonic. This allows for significant variation of the gas pressure across the layer. When this variation is taken into account the reconfinement point shifts further downstream compared to its position in the Komissarov-Falle (KF) model because the pressure in the layer turns out to be lower than the external one at the same distance from the source. For the case of a cold relativistic jet and uniform external gas, \cite{NS-09} found that the distance to the reconfinement point increases approximately by a factor of two.  A similar tendency was reported for hot relativistic jets in power law atmospheres with $\kappa>2$ \cite{KB-12}.    The underlying physical reason is relatively simple. In order to force the axial convergence of the initially expanding streamlines of a reconfined jet, the gas pressure across the shocked layer must be increasing outwards. Therefore at the reconfinement shock the pressure is lower than that in the external medium. The higher is the Mach number of the flow in the shocked layer, the stronger is the pressure difference. However, this qualitative argument has not yet been turned into a quantitative model that would rival the KF model in its simplicity and generality. 

In this paper, we describe a systematic study of the structure of steady-state unmagnetised jets undergoing reconfinement by the external gas pressure using computer simulations. The main goal of this study is to understand the reliability of KF model in predicting the position of the reconfinement point.  We focus on the setup most suitable to AGN jets with their moderate Lorentz factors and external atmospheres typical of their parent galaxies.  We start by presenting the KF model, including the new results required for comparison with our numerical models (Section \ref{sec:KF-model}).  The numerical method and the general features of the setup can be found in Section \ref{sec:method}.  The numerical solutions are described and analysed in Section \ref{sec:results}.  Section \ref{sec:discussion} discusses the implications of findings to the reconfinement of AGN jets and its possible connection to the Fanaroff-Riley division of extended extragalactic radio sources into two main groups. The conclusions are summarised in Section \ref{sec:conclusions}.

\section{KF model}  
\label{sec:KF-model}

Here we briefly summarise the KF model of reconfinement shocks in relativistic jets \citep{KF-97}.  This model utilises a number of simplifying assumptions. Its first  simplification is the axial symmetry.  It is also assumed that at the jet nozzle the flow speed, density and pressure are uniform and the velocity vectors define straight lines originating from a single point, the ``jet origin''. The initial half-opening angle of the jet $\theta_0 = r_0/z_0\ll1$, where $r_0$ is the nozzle radius and $z_0$ is its distance from the origin.  At the nozzle the jet is cold, $p_\ind{j}\ll\rho_\ind{j} c^2$ and relativistic $v_\ind{j}\approx c$. Hence, it is highly supersonic. Overall, these assumptions are designed  to represent a freely-expanding jet far away from its central engine where its plasma has adiabatically cooled. Although the uniformity is likely an oversimplification, given the absence of strong evidence in favour of any other particular jet structure, there is not much sense in choosing anything more complicated.   

The key equation of the model is a first order ODE which determines the shape function of the reconfinement shock, $r=r(z)$, where $r$ is the shock cylindrical radius at the distance $z$ from the jet origin. The natural initial condition for the function is $r(z_0)=r_0$; the shock curve originates from the edge of the nozzle.  In most astrophysical applications, the nozzle setup is not particularly meaningful and one would prefer to deal with a jet originating from the origin instead.  However, it makes perfect sense for simulated jets which do emerge from a nozzle of finite size.    Moreover, the astrophysical  solutions are easily recovered via the limit  $z_0 \to 0$ (and hence $r_0\to 0$).  
          
Finally, it is assumed that the angle between the shock curve and the $z$ axis is small, $dr/dz\ll 1$ and the pressure of the shocked jet gas matches that of the external gas at the same distance, $p_\ind{ext}(z)$.    Under these conditions the shock-shape equation reduces to  
\begin{equation}
 \dv{\theta}{z}  = - \theta_0 \sqrt{\frac{p_\ind{ext}(z)}{K}}\ , 
 \label{eq:S-ODE}
\end{equation}
where $\theta=r/z$, 
\begin{equation}
K=\frac{\mu L v_\ind{j}}{\pi c^2} \ ,
\label{eq:K}
\end{equation} 
where $\mu\approx17/24$,  $v$ is the constant jet speed and 
\begin{equation}
   L  =  \rho_\ind{j} \Gamma_\ind{j}^2 c^2 v_\ind{j}  \pi r_0^{2} \ .
   \label{eq:L}
\end{equation}
is its kinetic power \cite{KF-97,PK-15}. In the rest of the paper, we simply put $v_\ind{j}=c$ and $\mu=17/24$ into these expressions.   

\subsection{Power-law atmosphere}

For a power-law atmosphere with $p_{\ind{ext}} =p_c (z/z_c)^{-\kappa}$, equation~(\ref{eq:S-ODE}) with the initial condition  $\theta(0)=\theta_0$ has the solution 
\begin{equation}
   \theta(z) /\theta_0 = 1+\frac{A}{\delta} \left( \fracp{z_0}{z_c}^\delta -  \fracp{z}{z_c}^\delta \right) \ ,
\label{eq:theta}
\end{equation} 
where $\delta= 1-\kappa/2$ and $A^2=p_c z_c^2 \pi c/\mu L$.  The position $z_\ind{r}$  of the reconfinement point is defined by 
the equation 
\begin{equation}
    \theta(z_\ind{r})=0 \ .
\label{eq:th0}
\end{equation} 

For $\kappa>2$, equation (\ref{eq:th0}) has a real solution only if 
\begin{equation}
    \fracp{z_0}{z_c}^\delta >  \frac{|\delta|}{A}  \ .
\label{eq:cond1}
\end{equation} 
In order to explain this condition we compare the radial component of the ram pressure at the edge of the nozzle
$$
    p_\ind{ram,r}(z_0)=\mu \rho \Gamma^2 c^2 \theta_0^2 = \frac {\mu L}{\pi c  z_0^2} \ ,
$$
with the external gas pressure 
$$
   p_\ind{ext}(z_0) = p_c \fracp{z_0}{z_c}^{-\kappa} \,. 
$$
Provided $p_\ind{ram,r}(z_0)<p_\ind{ext}(z_0)$, the shock will dive towards the jet axis straight away without any period of expansion. 
It is easy to verify that this condition can be written as  
\begin{equation}
  \fracp{z_0}{z_c}^\delta > \frac{1}{A} \ , 
  \label{eq:cond2}
\end{equation} 
which is almost the same as the condition (\ref{eq:cond1}).  Thus, when $\kappa>2$ the shock reaches the jet axis only if the jet is over-expanded 
already at the nozzle.  Otherwise the shock surface asymptotically approaches the opening angle 
\beq
\theta_{\ind{s},\infty} = \theta_0\left(1 + \frac{A}{\delta} \fracp{z_0}{z_c}^\delta \right) \ .
\label{eq:th_asympt}
\eeq

For $\kappa<2$, equation (\ref{eq:th0}) always has a real solution, namely
\begin{equation}
    \fracp{z_\ind{r}}{z_c}^\delta=\frac{\delta}{A} \left(1+\fracp{z_0}{z_c}^\delta \frac{A}{\delta} \right) \ .
\label{recon}
\end{equation} 
For $z_\ind{r} \gg z_0$ one can ignore the second term in equation~(\ref{recon}) and obtain the asymptotic solution  
\begin{equation}
    \fracp{z_\ind{r}}{z_c} \simeq \delta^{1/\delta} \left( \frac{\mu L}{p_c z_c^2 \pi c} \right)^{1/2\delta} \ .
\label{zr-asymp}
\end{equation}
In units appropriate for AGN, this equation reads as 
\begin{equation}
    \log_{10} z_\ind{r,kpc} \simeq   \frac{1}{2\delta}\log_{10}\fracp{L_{44}}{p_{-9}} + u_\ind{KF}(\delta)  \ ,
\label{zrform-agn}
\end{equation}
where we put $z_c=1\ $kpc,  $p_c = 10^{-9}p_{-9}\,\mbox{dyn}/\mbox{cm}^2$,  $L=10^{44}L_{44}\,\mbox{erg}/\mbox{s}$  and 
\begin{equation}
    u_\ind{KF}(\delta) = \frac{1}{\delta} (\log_{10}\delta - 0.55 ) \ .
\label{uKF}
\end{equation}

In the marginal case $\kappa=2$, the solution to equation~(\ref{eq:S-ODE}) is 
\beq
    \theta(z)/\theta_0 = 1+ {A} \ln\frac{z_0}{z} \ .
\label{eq:theta-marg}
\eeq
Its reconfinement point is located at 
\beq
    z_\ind{r} = z_\ind{0}  \exp \fracp{1}{A}  \ .
\label{eq:zr-marg}
\eeq

\subsection{Arbitrary atmosphere}

Consider an external atmosphere with the gas pressure 
\beq
  p_\ind{ext} = p_c f(z/z_c) \, .
\eeq
In this case, equation (\ref{eq:S-ODE}) reads 
\beq
    \oder{\theta}{x} = -\theta_0 A \sqrt{f(x)} \ ,
\eeq
where $x=z/z_c$. Integrating this equation and applying the initial condition $\theta(z_0)=\theta_0$ one finds
\beq
\theta(z)/\theta_0= 1 - A \int\limits_{z_0/z_c}^{z/z_c} \sqrt{f(x)} dx \ , 
\eeq
which gives the reconfinement point as the solution of the integral equation 
\beq
   \int\limits^{z_r/z_c}_{z_0/z_c} \sqrt{f(x)} dx = \frac{1}{A} \,.
\label{FM-general}
\eeq

\section{Numerical Method and Simulation Setup}  
\label{sec:method}

In our study we construct numerical steady-state solutions for axisymmetric jets using the approach developed in \cite{KPL-15}. Here we briefly describe this unusual approach and refer the reader to the original paper for full details and test simulations.  

The approach is based on the close similarity between the axisymmetric two-dimensional (2D) steady-state equations  and axisymmetric one-dimensional (1D) time dependent equations of ideal relativistic magnetohydrodynamics. This similarity allows to approximate steady-state 2D solutions with the corresponding 1D time-dependent solutions upon the substitution $z=ct$. The approximation is reasonably accurate when applied 
to highly-collimated flows with the axial velocity $v_\ind{z}\approx c$.  The error of the approximation is of the order $v_\ind{r}/v_\ind{z} \simeq r/z$, which is about $10\%$ for a jet with the half-opening angle $\theta = 0.1$. This is acceptable  given the very large variations of the AGN jet parameters as well as the errors and uncertainties involved in their estimation from the observations. 
The main advantage of this approach is in its practicality and efficiency. One does not need to have a highly-specialised computer code designed to integrate the steady-state equations. Instead one can use one of many  codes designed to integrate time-dependent equations which are now readily available as open source.  Moreover, 
only the 1D equations have to be integrated. This differs from the traditional relaxation approach where the time-dependent 2D equations are integrated in a search for 2D steady-state solutions \citep[e.g.][]{UKRCL-99,kvkb-09,TMN-08}.  The latter approach is computationally more expensive and may fail completely when the sought steady-state solution is unstable.  
 
Our approach is similar to the traditional marching methods for supersonic flows where the steady-state solution at cross-section with $z_{n+1}$ is reconstructed using the information about the solution at $z_j$, $j\le n$
\citep[e.g.][]{DM-88,WF-85,wilson-87}. The difference is that we integrate forward in time and the solution at $t_{n}$ approximates the 2D steady-state solution at the distance $ct_{n}$ from the nozzle. The boundary conditions at the nozzle of the 2D steady-state problem correspond to the initial conditions for the 1D time-dependent problem.    

The 1D simulations were carried out with the Godunov-type code described in \cite{ssk-godun99}.  Although the approach can be used to study magnetised jets \cite{KPL-15}, here we limit ourselves to the hydrodynamic case. 
The scheme is based on a linear Riemann solver, it is second order accurate for smooth solutions and first-order accurate for solutions with discontinuities. Since the reconfinement problem involves a strong shock, it is the latter accuracy which is relevant here.   The code was run with the Courant number $C=0.4$.

\begin{figure}
\centering
\includegraphics[width=0.4\columnwidth]{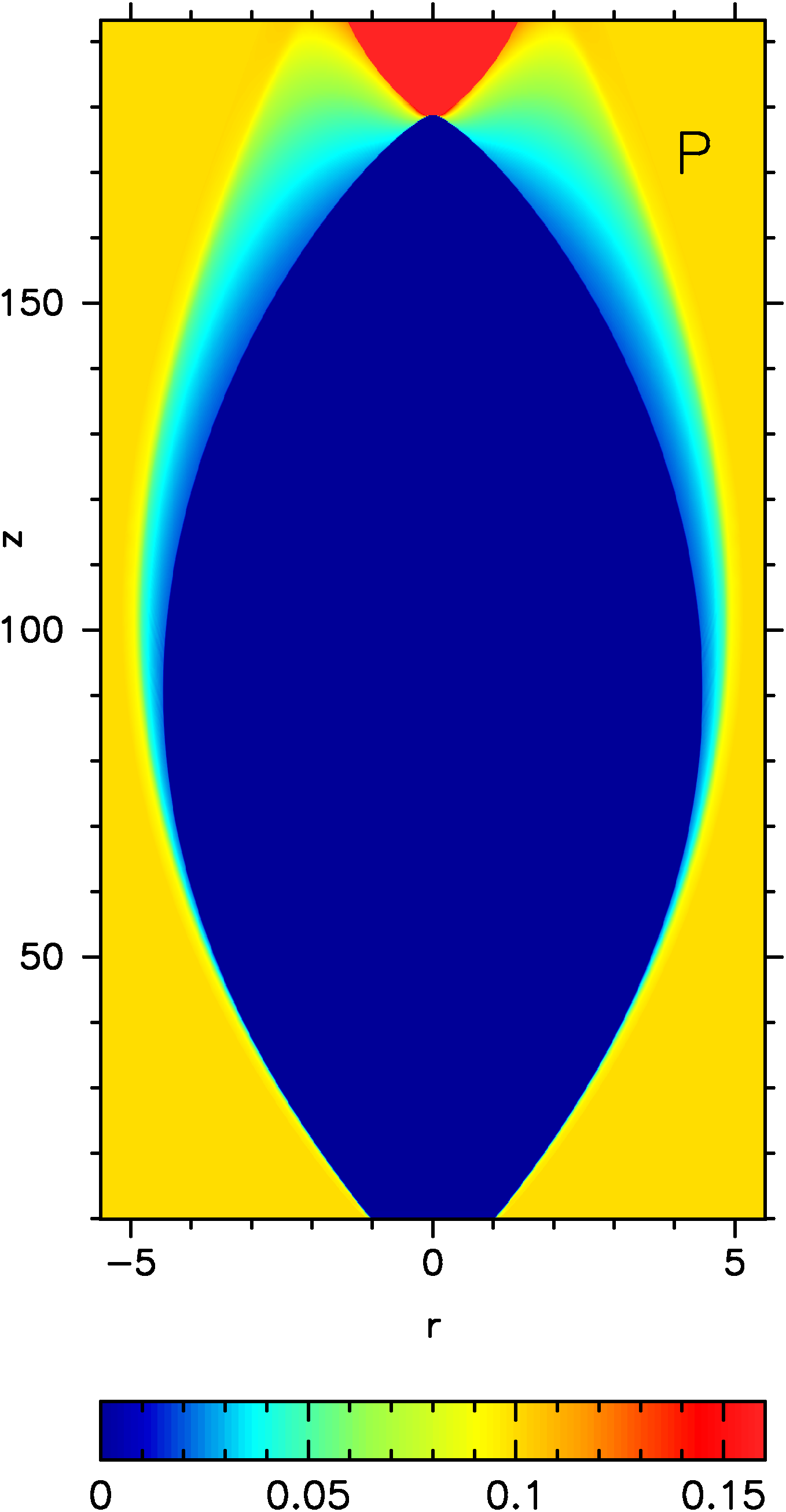}
\includegraphics[width=0.4\columnwidth]{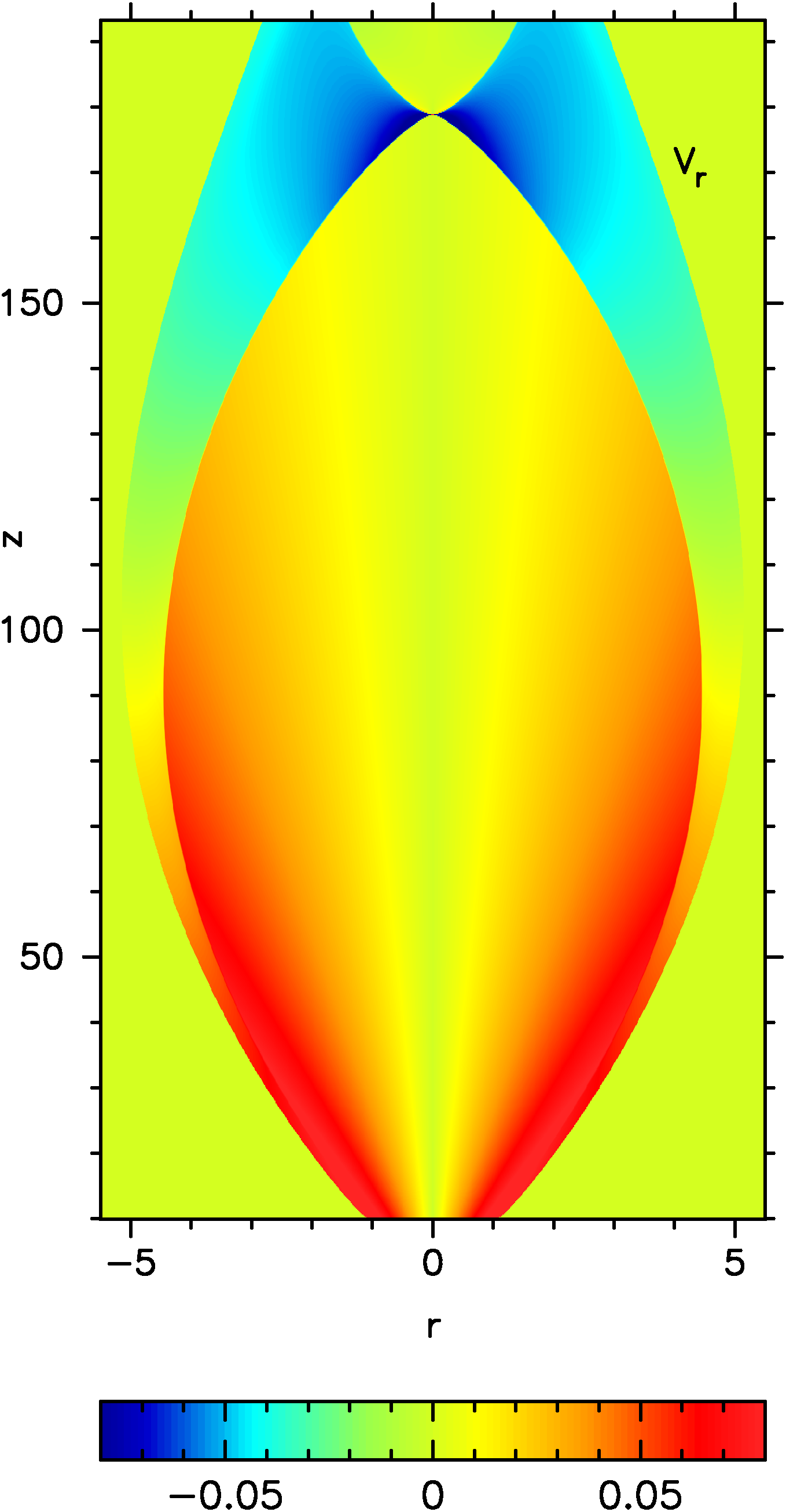}
\caption{A typical structure of axisymmetric steady-state jets undergoing reconfinement by the external gas pressure.  
The plots show the distributions of the gas pressure $p$ (left), and the radial velocity component $v_\ind{r}$ (right) for the case of uniform external pressure ( $\kappa=0$ ) and the initial jet density $\rho_\ind{j,0}=10$. The slight increase of the pressure near the symmetry axis is a numerical artifact. Note the different scaling in the vertical direction compared to the horizontal one.}
\label{fig:example}
\end{figure}

As explained in~\cite{KPL-15}, there are subtleties to this approach which one must take care with. In particular, the external density distribution plays no role in shaping the steady state jet solution but it matters for the time-dependent 1D solution. Indeed, as the 1D jet expands it drives waves into the external gas and hence loses energy.  In order to minimise these losses the external gas density has to be kept significantly below the jet density, causing the external sound speed to become relativistic and inhibiting the role of the external gas inertia. It must be emphasised that this is not physically motivated (since observations suggest the external density is orders of magnitude higher than than the jet density), but simply required for the successful conversion of the time-dependent code to one approximating the steady state case. 

In all simulations presented here, we utilise the same initial solution with the same parameters except the uniform initial jet density  $\rho_\ind{j,0}$, which is a proxy for the jet kinetic luminosity. The initial dimensionless jet radius is set to $r_0=1$. The external medium is static with $\rho_\ind{ext,0}c^2/p_\ind{ext,0}=10^{-3}$. The initial jet pressure is set to $p_\ind{j,0}=0.01p_\ind{ext,0}$. Its velocity field corresponds to a conical flow with the Lorentz factor $\Gamma_0=10$ and the half-opening angle $\theta_0=1/\Gamma_0=0.1$.  Hence the velocity vector 
$(v_r,v_z,v_\phi) \propto (\cos\theta,\sin\theta,0)$, where $\tan\theta= (r/r_0) \tan\theta_{j}$. When introducing time variation of the external pressure (and density) we assume that the nozzle is located at the dimensionless distance $z_0=10 r_0$ from the centre of the external gas distribution (e.g. the centre of an AGN).  The actual dimensionless value of $p_\ind{ext,0}$ at the nozzle is not important, for definiteness we put $p_\ind{ext,0}=0.1$ in the simulations described in Sections \ref{sec:power-law} and \ref{sec:king}, and  $p_\ind{ext,0}=1.0$ in the simulations of Section \ref{sec:mking} (The dimensionless value of the speed of light in the code is $c=1$.).  In all simulations, we use the relativistic EOS $w=\rho c^2 + \gamma p/(\gamma-1)$ with the ratio of specific heats $ \gamma=5/3$.  

In the KF model the geometry of the reconfinement shock does not depend on the equation of state of the jet plasma. When the inner structure of the shocked outer layer becomes important so may the equation of state.  Since the reconfinement shock is normally very oblique, only a small fraction of jet kinetic energy is converted into heat. Moreover at the nozzle the jet is cold and has only a rather moderate Lorentz factor.  Hence the shocked jet plasma cannot be ultra-relativistic and one would expect  the ratio of specific heats to be somewhere in the range  $ 4/3<\gamma<5/3$.  To test the sensitivity of the solutions to $\gamma$ we run models with $\gamma=4/3$ as well (see Appendix \ref{sec:eos}). The results show about 10\% variation of the reconfinement distance, which is not significant within the context of the problem. 

In the simulations we used a uniform computational grid with the basic resolution of  100 cells per initial jet radius. 
In our convergency study we verified that the numerical error of reconfinement distance never exceeds few percent and hence stays below the errors associated with the approximations of the approach. 

Figure~\ref{fig:example} illustrates the typical outcome of a single simulation run.   In this particular case the external atmosphere is uniform, $\kappa=0$, and the initial jet density   $\rho_\ind{j,0}=10^{2}p_\ind{ext,0}=10$. The corresponding dimensionless jet power is $L=3.1\times10^3$. 
The reconfinement shock is clearly visible in the pressure plot where it makes the boundary of the inner region of low pressure coloured in deep blue.  The shock can also be easily traced in the plot of the radial velocity which also helps to pinpoint the contact interface between the jet and the external gas.  The pressure plot also nicely illustrates the significant pressure gradient across the 
shocked outer layer of the jet, seen particularly well downstream of the point of the largest lateral expansion of the jet. 
The distance to the reconfinement point, measured via visual inspection of the plots zoomed onto the region where the shock approaches the jet axis, is $z_{\ind{r}} \sim 179$. 

Each such run determines one point of the $z_r$--$L$ diagram for the selected model of the external pressure. The collected data is presented in the form of tables in Appendix \ref{Data} and in the graphic form throughout the rest of paper.

\begin{figure*}
 \includegraphics[width=0.9\columnwidth]{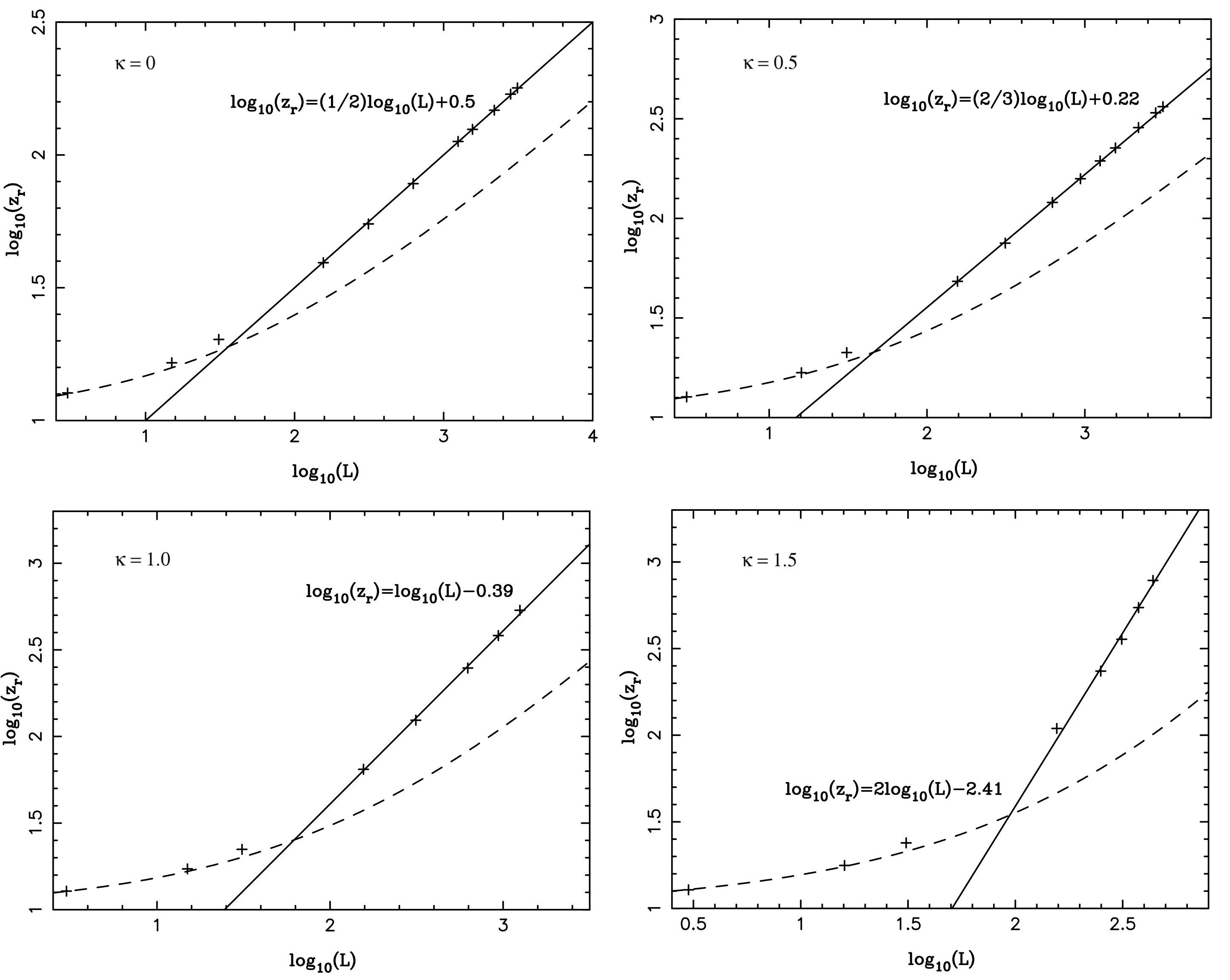}
\caption{The distance to the reconfinement point $z_{\ind{r}}$ as a function of the jet kinetic luminosity  $L$ in the case of power-law atmospheres with $\kappa=0.0$ (top left), $\kappa=0.5$ (top right), $\kappa=1.0$ (bottom left), $\kappa=1.5$ (bottom right).  In each panel, the crosses show the numerical results, the solid line shows the fitted high-$L$ power-law asymptote (also described by the panel's equation) and the dashed line shows the KF model solution as given by equation (\ref{recon}).}
\label{kappasim}
\end{figure*}

\section{Results}
\label{sec:results}

\subsection{Power-law atmosphere}
\label{sec:power-law}

We start by dealing with the simple case of power-law atmospheres. This case is important not only because 
it is described by simple analytic solutions in the KF model but also because it provides a reasonably good approximation for the conditions typical for the environment of astrophysical jets. Moreover, more realistic atmospheres can be approximated with piece-wise power-law functions. 

\subsubsection{The case $\kappa<2$}

We explored this case by studying models with $\kappa=0$, 0.5, 1.0, and 1.5. The key results are presented in Figure \ref{kappasim}  which shows the location of the reconfinement point as a function of the jet kinetic luminosity and compares it to the prediction of the KF model (see eq.\ref{recon}).  Apparently, the numerical results agree very well with the model at the low luminosities where the reconfinement point is close to the nozzle (see also Tables \ref{tabkappa0}-\ref{tabkappa2}).  This is because in this limit the reconfinement shock is not highly oblique and the post-shock Mach number of the flow is relatively low.  As the result, the pressure variation across the shocked layer is rather modest, as assumed in the KF model.   

For larger $L$, the shock obliqueness increases and so does the pressure variation. This leads to a significant deviation from the KF model.  However even in this regime, the flow eventually approaches the same power-law dependence of  $z_\ind{r}$ on $L$ as in the KF model (see Eq.\ref{zr-asymp}), but shifted upwards.    
 
For $z_\ind{r}\gg z_0$, equation (\ref{zr-asymp}) of the KF model predicts 
\beq
  \log_{10} z_r = \frac{1}{2-\kappa} \log_{10} L + s_\indt{KF}(\kappa) \ , 
  \label{eq:theory-predict}
\eeq
where 
\beq
s_\indt{KF}(\kappa) = \frac{2}{2-\kappa} \log_{10} \fracp{2-\kappa}{2} + \frac{1}{2-\kappa} 
\log_{10} \fracp{\mu}{\pi p_c z_c^\kappa} \ .
\label{eq:s_KF}
\eeq
The latter yields the values $s_\indt{KF}(0)=0.18$, $s_\indt{KF}(0.5)=-0.26$, $s_\indt{KF}(1.0)=-1.25$ and $s_\indt{KF}(1.5)=-4.70$.  In contrast, the numerical results are better described by 
\beq
  \log_{10} z_r = \frac{1}{2-\kappa} \log_{10} L + s(\kappa) \ , 
  \label{eq:simul-show}
\eeq
where $s(0)=0.50$, $s(0.5)=0.22$, $s(1.0)=-0.39$ and $s(1.5)=-2.41$ (these values are found via manual fitting).  Thus for the uniform external gas the reconfinement point is about twice as further out compared to the KF model. This is what was found earlier by \cite{NS-09}. Moreover, our results show that for $\kappa>0$ the discrepancy is even stronger, exceeding two orders of magnitude for the power-law atmosphere with $\kappa=1.5$.    
\cite{NS-09} showed that the difference between the KF model and the numerical results for the $\kappa=0$ case is related to the transverse distribution of pressure in the shocked layer at the jet boundary -- while the KF model assumes that it is constant an equal to the external pressure, in the numerical solutions  the pressure is variable and systematically lower than in the external gas. As a result, the numerical shock is weaker and more oblique than in the KF model, leading to a longer distance to RP. This is assumed to be the cause of the discrepancy for $\kappa>0$ also.
   
In dimensional form equation (\ref{eq:simul-show}) reads as 
\begin{equation}
\log_{10}(z_\ind{r}/l_\ind{u})=\frac{1}{2-\kappa} \log_{10}(L/p_\ind{u}l_\ind{u}^2c) + s(\kappa)\ ,
\label{kappa0best2}
\end{equation}
where $l_\ind{u}$ is the unit length (and hence $t_\ind{u}=l_\ind{u}/c$ is the corresponding unit time) and 
$p_\ind{u}$ is the unit pressure.  
Based on the typical parameters of the AGN problem, we put $l_\ind{u}=1 \ind{kpc}$, use such a unit of pressure that 
at $z=1 \ind{kpc}$ the external pressure $p_\ind{ext}=10^{-9}p_{-9} \,\ind{dyn}/\ind{cm}^2$ and measure the jet power in the units of $10^{44}\,\ind{erg}/\ind{s}$.  This leads to 
\begin{equation}
\log_{10} (z_\ind{r,kpc}) = \frac{1}{2-\kappa} \log_{10}\left(\frac{L_{44}}{p_{-9}}\right) +u(\kappa)\ ,
\label{kappa0bestdim}
\end{equation}
where $z_\ind{r,kpc}=z_\ind{r}/\mbox{kpc}$, $L_{44}=L/10^{44} \,\ind{erg}/\ind{s}$. As to the function $u(\kappa)$, its values at the four explored values of $\kappa$ are  $u(0)=-0.23$, $u(0.5)=-0.42$, $u(1)=-0.85$, $u(1.5)=-2.33$.   Motivated by the form of $s_\indt{KF}(\kappa)$, we adopt the approximation  
\begin{equation}
u(\kappa) \approx u_\ind{a}(\kappa)=\frac{2}{2-\kappa} \left(u_0+u_1 \log_{10}\frac{2-\kappa}{2}\right)\ ,
\label{eq:uk}
\end{equation}
where $u_{0}$ and $u_{1}$ are free parameters.  Imposing the conditions $u_\ind{a}(0)=u(0)$ and $u(1.5)=u_\ind{a}(1.5)$, we find $u_0=-0.23$ and $u_1=0.59$ and hence 
\begin{equation}
u(\kappa)\approx \frac{2}{2-\kappa} \left(-0.23+0.59 \log_{10} \frac{2-\kappa}{2}\right)\ .
\label{u}
\end{equation}
At $\kappa=0.5$ and $\kappa=1.0$ we find the approximation error to be around of $5\%$ and conclude that equations (\ref{kappa0bestdim}) 
and (\ref{u})  providing an accurate analytic representation of our numerical data for power-law atmospheres with $0<\kappa<2$.

\subsubsection{The case $\kappa=2$}

Using equation (\ref{eq:theta-marg}) one can easily verify that for the initial parameters used in our simulations the KF model predicts

\begin{equation}
\log_{10}{z_\ind{r}} = 1+ 0.065 \sqrt{L}\ .
\label{k2log}
\end{equation}

\begin{figure}
 \includegraphics[width=0.6\columnwidth]{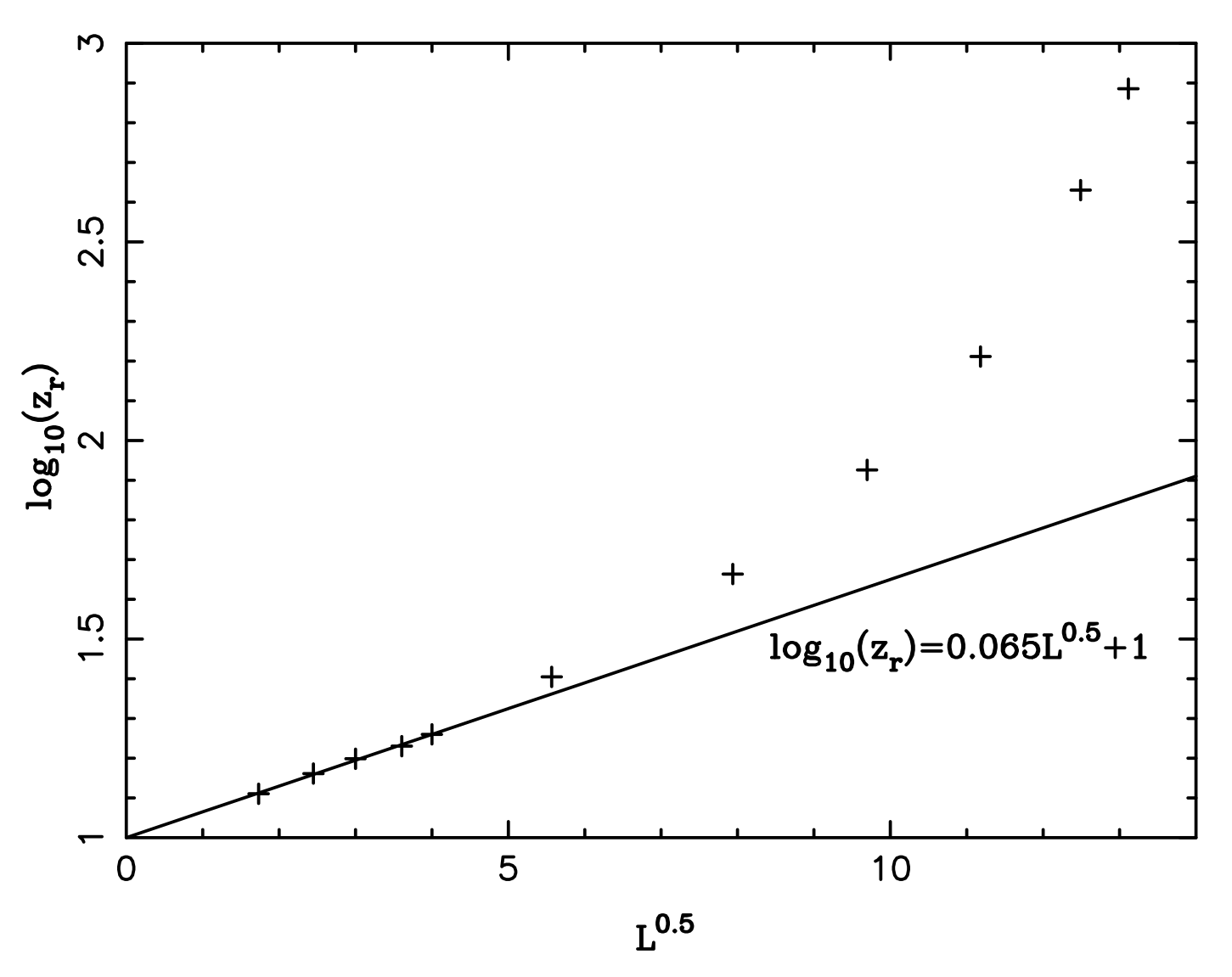}
\caption{The distance to the reconfinement point $z_{\ind{r}}$ against the jet kinetic luminosity  $L$ for the power-law atmosphere with $\kappa=2.0$. The solid line shows the prediction of the KF model.}
\label{kappa2sim}
\end{figure}

In contrast to the case $\kappa<2$, the initial jet radius enters the equation only via the jet kinetic luminosity and hence in the KF model it applies equally well both for low and high $L$. However, and in agreement with the $\kappa<2$ case, the numerical data agree with this prediction only when  $L$ is low enough to ensure that $z_\ind{r}$ is close to $z_0$ (see Figure~\ref{kappa2sim}). For larger $L$ the KF model still underestimates the reconfinement distance.  By analogy with the $\kappa<2$ case, one would expect the numerical  solution to eventually approach an asymptotic which runs parallel and above the line predicted by equation (\ref{k2log}).   However, Figure~\ref{kappa2sim} shows no indication of such a transition.

\subsection{King Atmosphere} 
\label{sec:king}

In this section we study the $z_\ind{r}$-$L$ relationship for the case when the external gas pressure is given by the function 
 \begin{equation}
p_{\ind{ext}}=p_\ind{c} \left[1+\fracp{z}{z_{\ind{c}}}^{2}\right]^{-\frac{\kappa}{2}}\ .
\label{eqking}
\end{equation}
Here $p_{c}$ is the pressure at $z=0$ and $z_{\ind{c}}$ is a characteristic length scale, called the core radius.  
Inside the core, for $z\ll z_{\ind{c}}$, the pressure profile is approximately uniform, $p_{\ind{ext}}=p_\ind{c}$, whereas outside of the core, for $z \gg z_{\ind{c}}$, it is approaches the power-law 
$$
p_{\ind{ext}}=p_\ind{c} \fracp{z}{z_{\ind{c}}}^{-\kappa} \, .
$$
The distribution (\ref{eqking}) is of the same form as in the King model for the density stellar and galactic clusters \cite{King1972}. It turns out that 
this model fits well the X-ray observations of coronal gas in massive elliptical galaxies but with $1<\kappa<1.5$ \cite{Mathews}  rather than   $\kappa=3$ in the original King model. 

\begin{figure}
 \includegraphics[width=0.45\columnwidth]{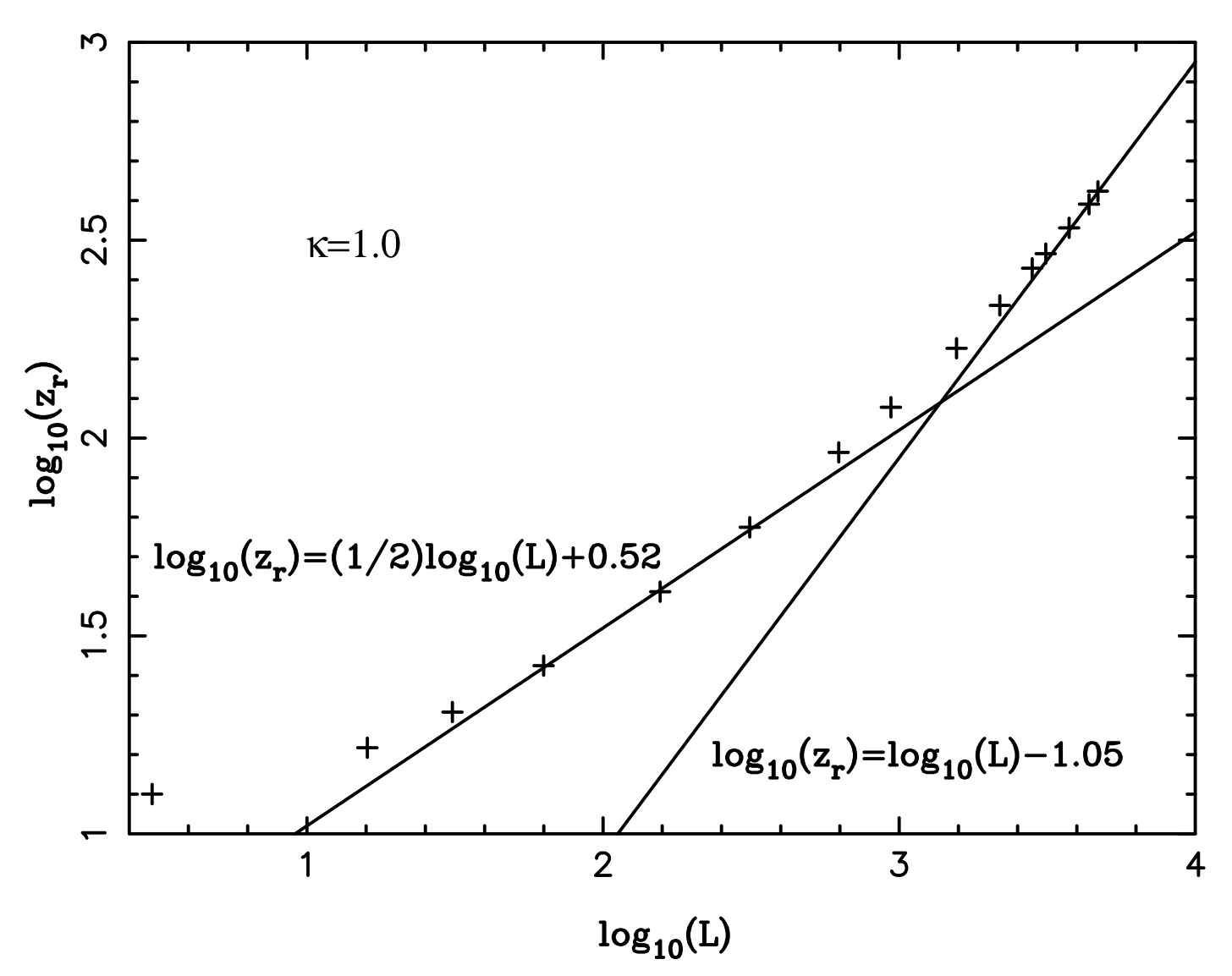}
\includegraphics[width=0.45\columnwidth]{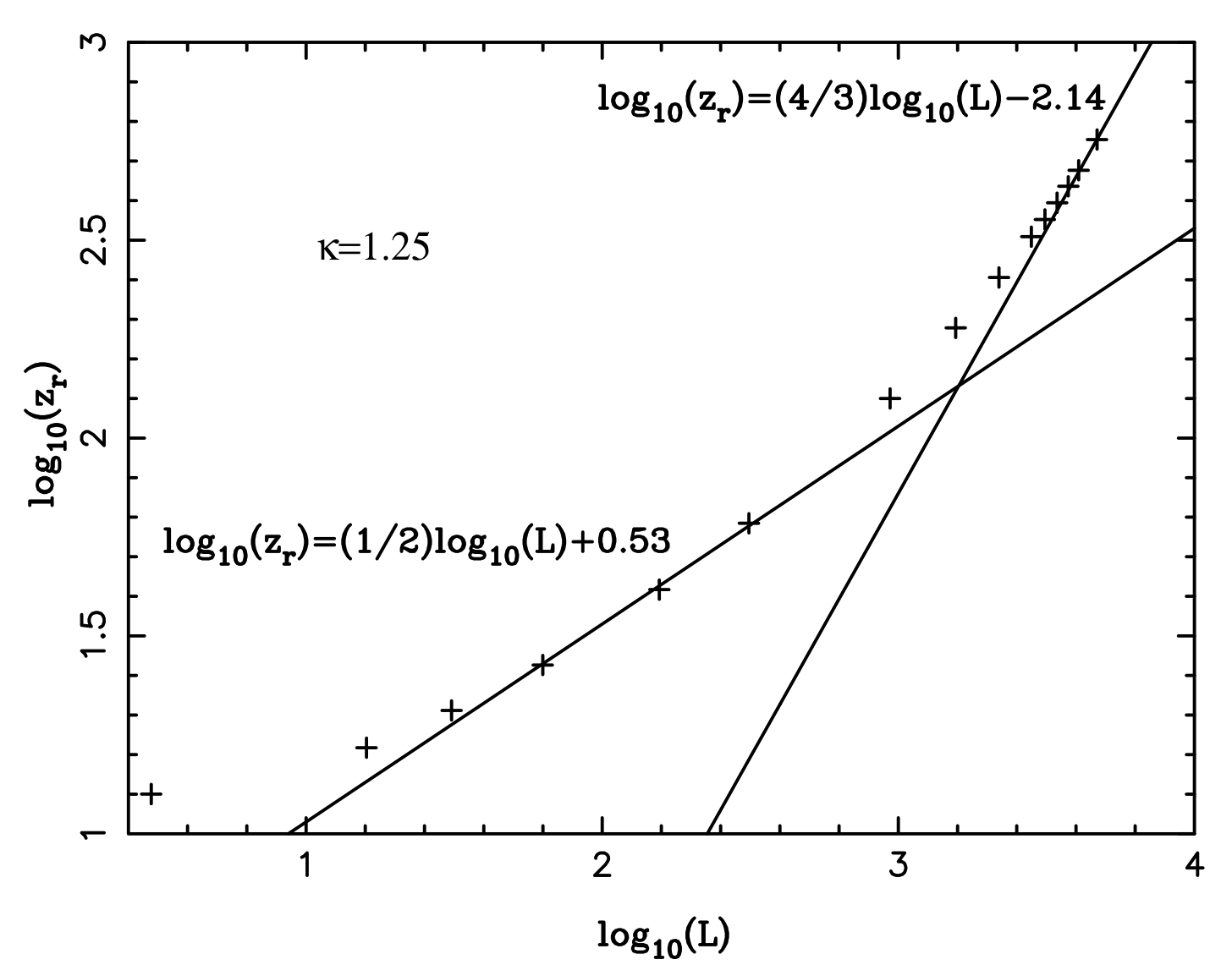}
\includegraphics[width=0.45\columnwidth]{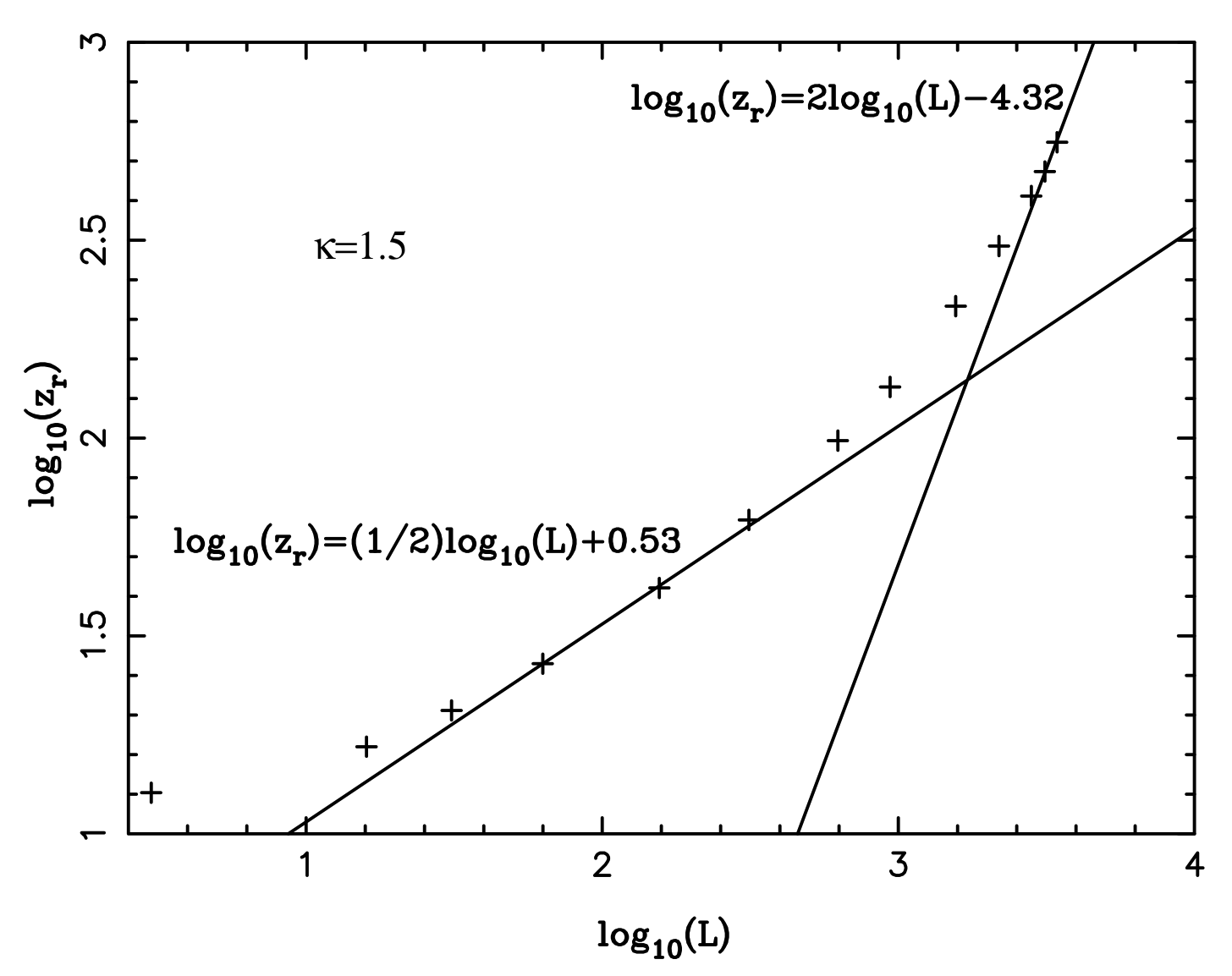}
           \caption{The distance to the reconfinement point $z_{\ind{r}}$ as a function of the jet kinetic 
luminosity  $L$ for King atmospheres with $\kappa=1$, $1.25$ and 
$1.5$.  In each panel, the crosses show the numerical results and the continuous lines show  their fitting by power-law asymptotes at low and high luminosity domains. Next to each asymptote its exact equation is given as well.}
 \label{zc50kingsim}
\end{figure}

In this study we put $z_{\ind{c}}=50$, which is five times the nozzle distance ($z_0=10$) and fifty times the nozzle radius ($r_0=1$). This allows us to study the effect of the change in the external pressure profile on the $z_\ind{r}$-$L$ relationship. Based on the results for power-law atmospheres, we expect to see $z_\ind{r} \propto \sqrt{L}$ when the reconfinement occurs inside the core and 
$z_\ind{r} \propto L^{1/(2-\kappa)}$ when the RP is well outside of the core. The normal resolution of these simulations is 100 cells per initial jet radius, with twice that resolution used to check the numerical convergence. 
Figure~\ref{zc50kingsim} shows the results for $\kappa=1.0$, 1.25 and 1.5, covering the whole range of the power index found in the observations of giant elliptical galaxies.

We start by analysing in detail the case $\kappa=1.0$.  As expected, for $z_\ind{r} \ll z_\ind{c}$ the results are almost the same as for the power law model with $\kappa=0$
\begin{equation}
\log_{10}(z_\ind{r})=\frac{1}{2}  \log_{10}(L)+0.50\ ,
\label{bestfit1}
\end{equation}
Indeed, as the jet is supersonic it is unaware of the steep decline in the external pressure at $z>z_\ind{c}$ until it enters this region. 
For $z_r\simeq z_0$ there is a significant deviation from this law, which is caused by the effects of the finite nozzle radius; just like in the simulations described in Section \ref{sec:power-law}.  Equation~(\ref{bestfit1}) applies equally well to the King models for the other two explored values of $\kappa$.  

For $z_\ind{r} \gg z_\ind{c}$, the numerical solution is well fitted by 
\begin{equation}
\log_{10}(z_{\ind{r}})=\log_{10}(L)-1.05\ ,
\label{bestfit2}
\end{equation}
thus confirming the anticipated $z_\ind{r} \propto {L}$ dependence. The numerical constant in equation~(\ref{bestfit2}) is lower than $-0.39$ found for the power-law atmosphere with the same 
$\kappa$ (see equation~\ref{eq:simul-show}).  This is because in both models the inlet external pressure is the same but in the King model the external pressure decreases slower with $z$ inside the core. As the result, 
outside of the core the pressure is higher than in the corresponding power-law model with the same $\kappa$, pushing the reconfinement shock inside the jet at a faster rate. 

We can then rewrite equations~(\ref{bestfit1}) and (\ref{bestfit2}) in the rescaled form by 
utilising the unit length $l_\ind{u}=1\,$kpc and  such a unit of pressure $p_u\ind{}$ that  $p_\ind{c}=10^{-9}p_{-9} \,\ind{dyn}/\ind{cm}^2$. Measuring the jet power in the units of $L_\ind{u}=10^{44}\,\ind{erg}/\ind{s}$, we obtain the rescaled versions of equations (\ref{bestfit1}) and (\ref{bestfit2})
\begin{equation}
\log_{10}\fracp{z_\ind{r}}{z_\ind{c}} = \frac{1}{2} \log_{10} (x)    - 0.23\ ,
\label{bestfit1dim}
\end{equation}
for $z_\ind{r} \ll z_\ind{c}$ and 
\begin{equation}
\log_{10}\fracp{z_\ind{r}}{z_\ind{c}}=\log_{10}(x) -0.81\ ,
\label{bestfit2dim}
\end{equation}
for $z_\ind{r} \gg z_\ind{c}$, where  
\begin{equation}
x=\frac{L_{44}}{p_{c,-9} z_{\ind{c,kpc}}^{2}} \ .
\label{x}
\end{equation}
For $z_\ind{r}/z_\ind{c}=1$ equation (\ref{bestfit1dim}) gives $x\approx 3$ and hence the above conditions on 
$z_\ind{r}$ can be translated into $x\ll 3$ and $x\gg3$ respectively. 

Repeating the process for the other two King models, we find that  
\begin{equation}
\log_{10}\fracp{z_\ind{r}}{z_\ind{c}} = \frac{1}{2} \log_{10} (x)    - 0.23\ ,
\label{king-comb-low}
\end{equation}
for $x \ll 3$ and
\begin{equation}
\log_{10}\fracp{z_\ind{r}}{z_\ind{c}}=\frac{1}{2-\kappa}\log_{10}(x) + u(\kappa)\ ,
\label{king-comb-high}
\end{equation}
for $x \gg 3$, where $u(1)=-0.81$, $u(1.25)=-1.25$ and $u(1.5)=-2.14$. (The asymptotes intersect at about the same point, which corresponds to $x\approx 10$ and $z_\ind{r}\approx 2 z_\ind{c}$.)

Finally, we derive the approximate formula applicable for $u(\kappa)$, $1<\kappa<1.5$. To this aim we use the same form for $u(\kappa$) as in equation~(\ref{eq:uk}) and demand that $u(1)=-0.81$ and $u(1.5)=-2.14$. This allows us to fix the parameters $u_0$ and $u_1$ 
and arrive to 
\begin{equation}
u(\kappa)=\frac{2}{2-\kappa} \left(0.28-0.43 \log_{10}{\frac{2-\kappa}{2}}\right)\ .
\label{eq:uk2}
\end{equation}
This function yields  $u(1.25)=-1.24$, which is very close to the value of  -1.25 obtain directly from the numerical data, implying the maximum error of the order of one percent.

\subsection{King atmosphere with central cusp}
\label{sec:mking}

It is all but impossible for the King model to hold at the very centre of a giant elliptical galaxy, where the gravitational pull of its central supermassive black hole becomes the dominant force. Inside the Bondi radius the force is expected to drive an accretion flow that ultimately feeds the central engine of its AGN. Strong winds and jets from AGN are other important factors that further complicate the picture. So far there are only handful of galaxies where the X-ray observations allow to explore the hot gas distribution inside the Bondi radius.  The data are consistent with a central cusp with the density $\rho \propto r^{-1}$ and a significantly slower temperature variation \cite{Wang-13,Wong-14,Russell-15}.

In this section, we focus on the effect such a cusp may have on the reconfinement scale of AGN jets.
Based on the results for M87 by \cite{Russell-15}, we now adopt the following model for galactic X-ray gas:      
\begin{equation}
p_\ind{ext}(z)= \left\{\begin{array}{lr}
            K z^{-1}, & \text{for }z\leq z_{\ind{cusp}}\\
           p_\ind{c}\left(1+ (z/z_\ind{c})^{2}\right)^{-{\kappa}/{2}}, & \text{for }z\geq z_{\ind{cusp}}
             \end{array}\right. \ ,
\label{Pmodking}
\end{equation}
where $z_{\ind{c}} \sim 10 z_{\ind{cusp}}$ and the constant $K$ is such that the distribution is continuous at the cusp radius $z_\ind{cusp}$.  

\begin{figure}
\includegraphics[width=0.6\columnwidth]{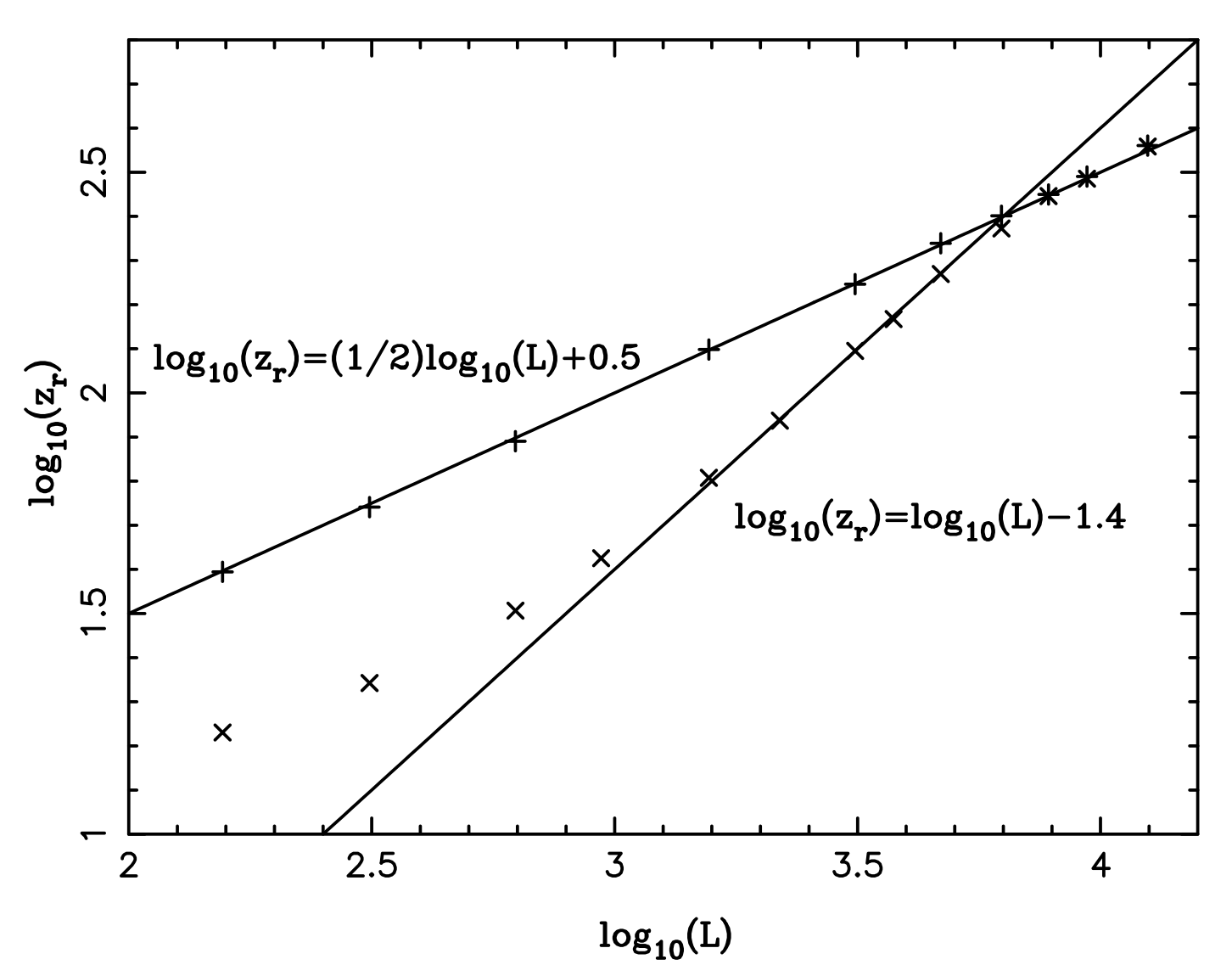}
\caption{ The distance to the reconfinement point as a function of the jet kinetic power for the King atmosphere with  ($\times$) and without ($+$) the central cusp.  The solid lines show the fitted power law asymptotes.  }
\label{zc1000modking}
\end{figure}

Like in all other simulations, in this model the jet nozzle is fixed to $z_0=10$. As to the other length scales, we put
$z_{\ind{cusp}}=100$ and  $z_{\ind{c}}=1000$ and focus of the deviation of the $z_\ind{r}$-$L$ relationship from that of the  original King profile. In the simulations we increased the value of the external pressure at the nozzle from $p_\ind{ext,0}=0.1$ to $p_\ind{ext,0}=1.0$ so that the value of $p_\ind{c}$ remains exactly the same as in the King profile introduced in  Section~\ref{sec:king}.

This model has three well-defined regions where the pressure distribution is close to a power law: 
(i) the cusp region $z<z_\ind{cusp}$, where $p_\ind{ext}\propto z^{-1}$; (ii) the core region $z_\ind{cusp}<z<z_\ind{c}$, where 
$p_\ind{ext}\approx$const;  (iii) the outer region $z>z_\ind{c}$ where $p_\ind{ext}\propto z^{-\kappa}$.  Within each of these regions one would expect the $z_\ind{r}-L$ dependence to be similar to that found for the pure power-law atmosphere with the same index.  
(i) $z_\ind{r} \propto L$  for the cusp (ii) $z_\ind{r} \propto \sqrt{L}$  for the core.  The results of our simulations are fully consistent with this expectation.

We find that inside the cusp the numerical solution gradually approaches the asymptote  
\begin{equation}
\log_{10}(z_\ind{r})=\log_{10}(L)-1.40
\label{modbestfit1}
\end{equation}
for $z_0 \ll z_\ind{r}<z_\ind{cusp}$.
This is the same power-law $z_\ind{r}\propto L$ as the one found for the pure power-law atmosphere $p_\ind{ext}\propto z^{-1}$ in Section  \ref{sec:power-law}. Note that the constant in the equation (\ref{modbestfit1}) differs from $s(1)=-0.39$ in equation (\ref{eq:simul-show}). This is simply because in the setup of this problem the external pressure at the nozzle  $p_\ind{ext,0}$ is increased from 0.1 to 1.0 which reduces the reconfinement scale ( as $z_\ind{r}\propto p_\ind{ext,0}^{-1}$ in the asymptotic regime).  The significant deviation from the asymptote near $z_0$ is again the effect of the non-vanishing nozzle radius.   

Outside of the cusp but still inside the core, the numerical solution is well approximated by the core asymptote found for 
the original King atmosphere without cusp   
\begin{equation}
\log_{10}(z_{\ind{r}})=\frac{1}{2} \log_{10}(L)+0.50
\label{modbestfit2}
\end{equation}
(see Sec.\ref{sec:king}). Thus the cusp has little effect on the reconfinement of jets which have $z_\ind{r} > z_\ind{cusp}$ in the original King model.  However, it can significantly reduce the reconfinement scale of the less powerful jets.

The transition from one asymptote to another occurs around $z\approx 2.5 z_{\ind{cusp}}$.  This transition is very sharp which seems to be reflecting the sharp change in the pressure gradient at $z_{\ind{cusp}}$ introduced by the 
equation~(\ref{Pmodking}).
 
With the scaling corresponding to the typical AGN parameters, these results read
\begin{equation}
\log_{10}\fracp{z_\ind{r}}{z_\ind{c}}  =\log_{10}(x) +0.14\ ,
\end{equation}
for  $x<0.2$ ($z<2.5 z_\ind{cusp}$) and
\begin{equation}
\log_{10} \fracp{z_\ind{r}}{z_\ind{c}}  = \frac{1}{2} \log_{10}(x) -0.23\ ,
\end{equation}
for $x>0.2$ ($z> 2.5z_\ind{cusp}$), where the luminosity parameter $x$ is the same as in equation~(\ref{x}).

\section{Discussion}
\label{sec:discussion}

As we have pointed out in the Introduction, the reconfinement of AGN jets can manifest itself via for a number  spectacular  phenomena. In particular, it may trigger various instabilities which could be behind the global division of extended extragalactic radio sources into two main groups, distinctive both in terms of the source luminosity and morphology \cite{FR-74}. The powerful FR-2 sources, of which Cyg A is the most popular example, display one-sided jets that can be traced all the way to the leading hot spots, the most remote and brightest features of their radio lobes. This morphology is consistent with the shock interaction between the external gas and collimated supersonic relativistic flows produced by AGN.  On the contrary, the jets of  FR-1 sources are less collimated, two-sided and appear to gradually dissolve inside the extended lobes \citep{RPFF90}. These lobes do not have leading hot spots (the radio galaxy 3C 31 is a good example). Such structure is more consistent with a transonic sub-relativistic turbulent jet being mixed with the lobe plasma, and interstellar gas, via boundary interactions.  Based on the energetics of cavities produced by the AGN jets in the hot gas surrounding many massive galaxies, their kinetic power ranges at least from $L_\ind{min}\approx 10^{42}\mbox{erg/s}$ to $L_\ind{max}\approx10^{46}\mbox{erg/s}$, with the FR division occurring roughly  at $L_{\ind{FR}}\approx 10^{44}\mbox{erg/s}$  \citep{Cavagnolo}. In fact, the critical power seems to depend on the mass of the parent galaxy \cite{OL-94}.

The typical thermal pressure of galactic coronas seems high enough to terminate the initial free expansion of less powerful AGN jets. According to the recent study \citep{PK-15}, weak FR-1 jets can be reconfined already inside the central cores of these coronas. On the contrary, the FR-2 jets are powerful enough to pass through the cores and remain free for much longer. Moreover, the critical power separating jets reconfined inside the core from those reconfined outside of it depends on the optical luminosity of the parent galaxy, in line with the observations \cite{OL-94}.  These correlations suggest that the jet reconfinement may well be one of the key physical processes behind the FR division. Moreover, the reconfinement restores the jet connectivity and hence allows development of global instabilities, which may explain why the FR-1 jet become turbulent on the kpc scales  \cite{F-91,PK-15,GK-18NatAs}.          

\begin{figure}
        \includegraphics[width=0.65\textwidth]{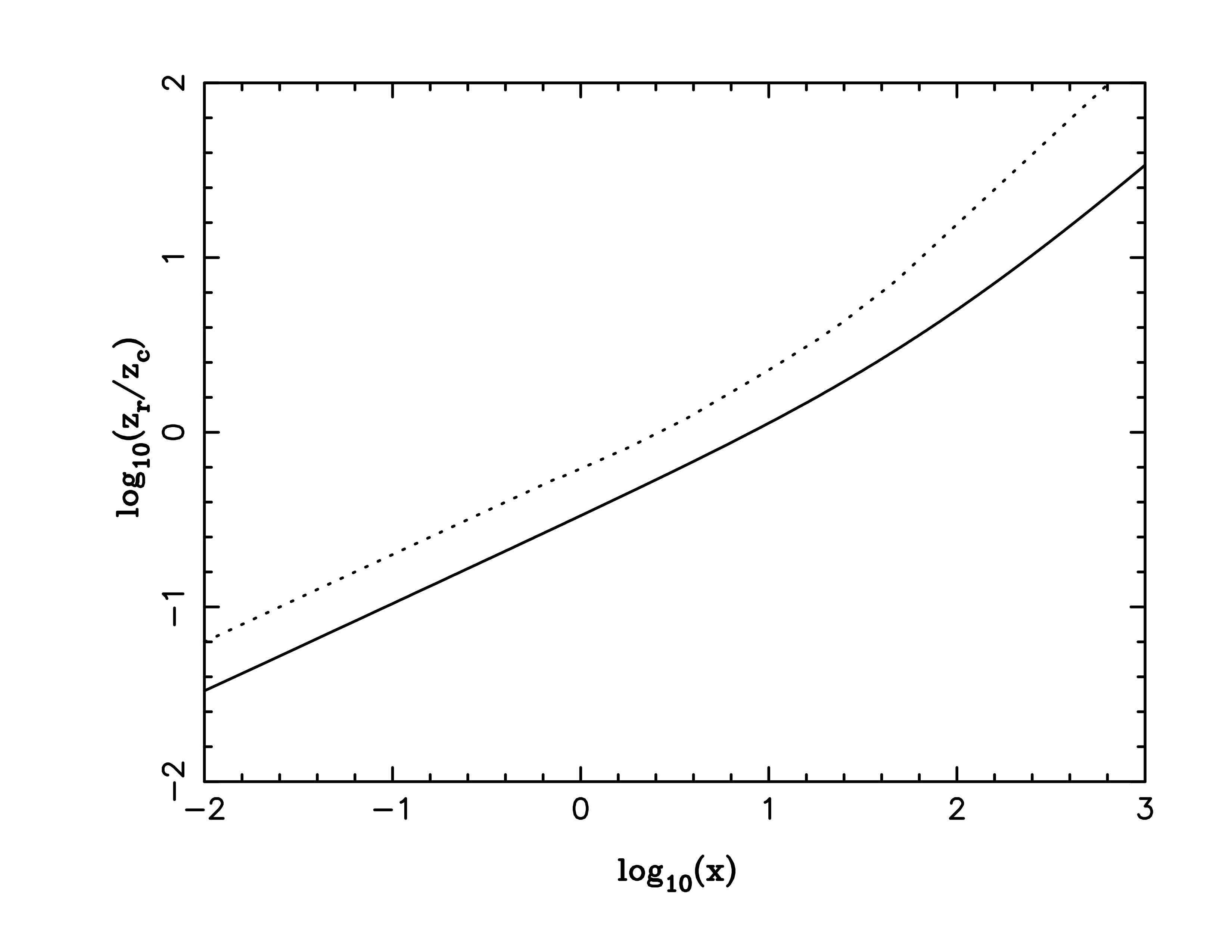}
        \includegraphics[width=0.65\textwidth]{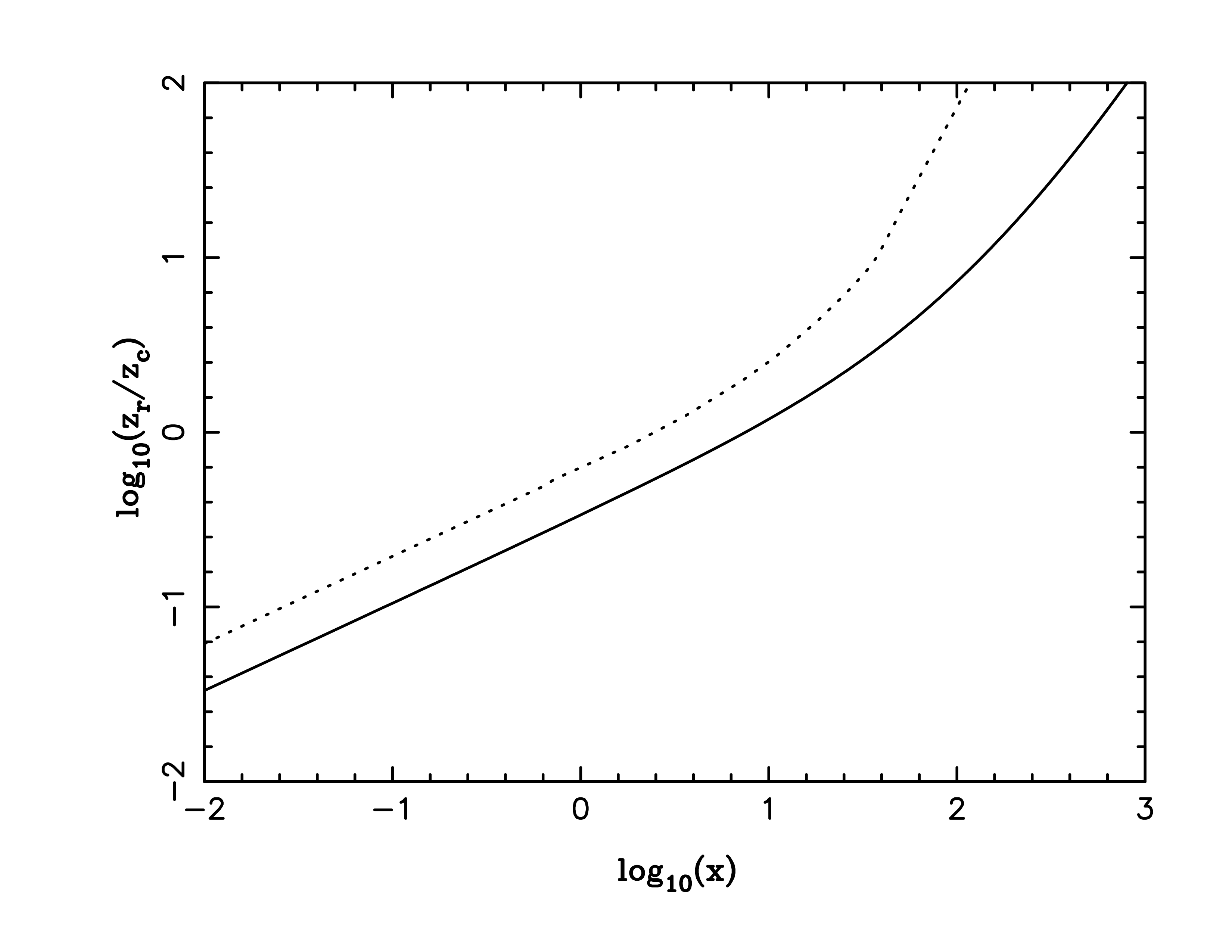}
        \caption{$\log_{10}(z_{\ind{r}}/z_{\ind{c}})$ against $\log_{10}x$, where $x=L_{44}/p_{c,-9} z_{\ind{c,kpc}}^{2}$ for the King 
        model of the external gas pressure with $\kappa=1$ (top) and $\kappa=1.5$. 
The solid line shows the predictions of the KF model, whilst the dotted line shows the corresponding solution based on our computer simulations.}
 \label{frdivision}
\end{figure}

The analysis of the jet reconfinement carried out in \citep{PK-15} is based on the KF model. Given the importance of the issue, we now assess the accuracy of their calculations.  Figure~\ref{frdivision} compares the $z_\ind{r}$-$L$ curves for the case of the King atmosphere obtained using equation~\ref{FM-general} of the KF model (solid lines) and our numerical approach (dotted lines).  As one can see, inside the core the difference between the results is rather small, with the reconfinement distance being about twice shorter in the KF model, in agreement with what has been found for a uniform distribution of external gas (Section~\ref{sec:power-law}).  Outside of the core the difference grows, in accordance with the results obtained for power-law atmospheres (Section~\ref{sec:power-law}).  In particular, for $\kappa=1.0$ the ratio of the reconfinement radii obtained numerically and using the KF model  asymptotically approaches $z_\ind{r}/z_\ind{r,KF}\approx 7$, the value found for the pure power-law atmosphere with this  $\kappa$.     For $\kappa=1.5$ the predicted asymptotic value of $z_\ind{r}/z_\ind{r,KF}$ exceeds two orders of magnitude. This explains why in Figure~\ref{frdivision}  the divergence between curves keeps increasing -- the asymptotic regime is not reached within the studied domain.     

If following \citep{PK-15} we identify the borderline between FR-1 and FR-2 sources with the jet reconfinement at the edge of the coronal core then the critical jet luminosity corresponds to the power parameter $x=3$ (see Sec.\ref{sec:king}).  This yields,    
\beq
     L_\ind{FR}  = 3 \times 10^{44}\, p_\ind{0,-9}\, z^2_\ind{c,kpc} \,\mbox{erg/s} \,,
     \label{LFR-corrected}
\eeq 
which is in a good agreement with the observations. As we have established in Section \ref{sec:mking}, central cusps have no effect on this division but reduce the reconfinement scale of jets with very low power.  For example the cusp found in M87 would effect the reconfinement of jets with the power parameter $x<0.1$.  The corresponding kinetic luminosity of such jets is below   
\beq
     L_\ind{css} =  10^{43}\, p_\ind{0,-9}\, z^2_\ind{c,kpc} \,\mbox{erg/s} \,.
     \label{L-css}
\eeq 
They could be identified with some of the so-called compact steep spectrum double radio sources \citep{FF90}.  

In 3D simulations, the reconfinement of unmagnetised jets is characterised by a rapid development of 
centrifugal instability and additional heating of the shocked jet plasma.  This may push the reconfinement 
point towards the jet source compared to its 2D steady-state location, by a factor two or so \citep{GK-18NatAs}. 
In principle, this may have some observational significance.

Falle \cite{F-91} developed an axisymmetric self-similar hydrodynamic model of FR-2 sources produced by initially free expanding conical jets. He demonstrated that these jets must become confined by the gas pressure of the hot plasma bubble (often called a cocoon or lobe) which they inflate when interacting with the preexisting interstellar, galactic and intergalactic gas. As the bubble pressure can be significantly higher than that of the galactic coronal gas, the reconfinement scale can be significantly lower compared to what we find in our study, particularly when the bubble is still located inside the galaxy.  The dynamic nature of this problem makes it more challenging and the outcome less certain. A number of recent studies have begun to address this issue \cite{MBR16,TB16} but much more remains to be done in this direction.

\section{Conclusion}
\label{sec:conclusions}

We have studied the reconfinement of initially free conical hydrodynamic jets by the external pressure with application to AGN jets. 
To this end we considered a few models of the external gas distribution relevant to this problem, including power-law atmospheres, the isothermal King model and the King model modified by the introduction of a central cusp region. Steady-state axisymmetric jet solutions were found by means of 1D time-dependent simulations as described in \citep{KPL-15}. Using this approach we constructed  more than one hundred jet models which allowed us to study the dependence of the reconfinement scale on the jet power  for each of the external gas models in details.  These results were compared to the simple semi-analytical model of reconfined jets by Komissarov \& Falle \cite{KF-97} based on the assumption of constant pressure across the shocked outer 
layer of such jets. 

In the case of power-law atmospheres with the pressure $p\propto z^{-\kappa}$ ($0\le\kappa\le 2$), the difference between the two numerical and KF solutions increases with $\kappa$. It is most pronounced in the asymptotic regime where the initial jet radius becomes a small parameter.  In fact this regime is most relevant to the AGN jet problem under consideration.  We have found that in the asymptotic regime the KF model underestimates the reconfinement distance by the factor of two, three, seven and two hundred (!) for $\kappa=0$, 0.5, 1.0 and 1.5 respectively.  For $\kappa=2$ the numerical solution is qualitatively different from KF solution.  The difference is attributed to the pressure gradient developing across the supersonic shocked outer layer of a reconfined jet, which is ignored in the KF model.  
     
For the King distribution of the galactic coronal gas our results also show some significant deviations from the predictions of the KF model.  In particular, we find that for the powerful jets, which are reconfined outside of the King core the reconfinement scale can exceed that given by the KF model by up to one or two orders of magnitude, depending on the power index of the envelope. However, for the jets reconfined inside the core (and outside of the central cusp) the difference with KF model is rather minor, with the reconfinement scale being longer only by a factor of two. Thus we confirm the conclusion by \cite{PK-15} that for the typical power of FR-1 jets the reconfinement takes place inside the King core and for the typical power of FR-2 jets well outside of it.   

The central cusp with $p\propto z^{-1}$ has a strong impact only on the jets which are reconfined inside the cusp. Their reconfinement scale can be significantly shorter compared to the one expected in the King model without the cusp.  

\section{Acknowledgments}
SSK was supported by STFC via grant No. ST/N000676/1. 

\bibliographystyle{bmc-mathphys}
\bibliography{zrl,mix,jets,komissarov,numerics}


\begin{thebibliography}{47}
\ifx \bisbn   \undefined \def \bisbn  #1{ISBN #1}\fi
\ifx \binits  \undefined \def \binits#1{#1}\fi
\ifx \bauthor  \undefined \def \bauthor#1{#1}\fi
\ifx \batitle  \undefined \def \batitle#1{#1}\fi
\ifx \bjtitle  \undefined \def \bjtitle#1{#1}\fi
\ifx \bvolume  \undefined \def \bvolume#1{\textbf{#1}}\fi
\ifx \byear  \undefined \def \byear#1{#1}\fi
\ifx \bissue  \undefined \def \bissue#1{#1}\fi
\ifx \bfpage  \undefined \def \bfpage#1{#1}\fi
\ifx \blpage  \undefined \def \blpage #1{#1}\fi
\ifx \burl  \undefined \def \burl#1{\textsf{#1}}\fi
\ifx \doiurl  \undefined \def \doiurl#1{\textsf{#1}}\fi
\ifx \betal  \undefined \def \betal{\textit{et al.}}\fi
\ifx \binstitute  \undefined \def \binstitute#1{#1}\fi
\ifx \binstitutionaled  \undefined \def \binstitutionaled#1{#1}\fi
\ifx \bctitle  \undefined \def \bctitle#1{#1}\fi
\ifx \beditor  \undefined \def \beditor#1{#1}\fi
\ifx \bpublisher  \undefined \def \bpublisher#1{#1}\fi
\ifx \bbtitle  \undefined \def \bbtitle#1{#1}\fi
\ifx \bedition  \undefined \def \bedition#1{#1}\fi
\ifx \bseriesno  \undefined \def \bseriesno#1{#1}\fi
\ifx \blocation  \undefined \def \blocation#1{#1}\fi
\ifx \bsertitle  \undefined \def \bsertitle#1{#1}\fi
\ifx \bsnm \undefined \def \bsnm#1{#1}\fi
\ifx \bsuffix \undefined \def \bsuffix#1{#1}\fi
\ifx \bparticle \undefined \def \bparticle#1{#1}\fi
\ifx \barticle \undefined \def \barticle#1{#1}\fi
\ifx \bconfdate \undefined \def \bconfdate #1{#1}\fi
\ifx \botherref \undefined \def \botherref #1{#1}\fi
\ifx \url \undefined \def \url#1{\textsf{#1}}\fi
\ifx \bchapter \undefined \def \bchapter#1{#1}\fi
\ifx \bbook \undefined \def \bbook#1{#1}\fi
\ifx \bcomment \undefined \def \bcomment#1{#1}\fi
\ifx \oauthor \undefined \def \oauthor#1{#1}\fi
\ifx \citeauthoryear \undefined \def \citeauthoryear#1{#1}\fi
\ifx \endbibitem  \undefined \def \endbibitem {}\fi
\ifx \bconflocation  \undefined \def \bconflocation#1{#1}\fi
\ifx \arxivurl  \undefined \def \arxivurl#1{\textsf{#1}}\fi
\csname PreBibitemsHook\endcsname

\bibitem{KF-97}
\begin{barticle}
\bauthor{\bsnm{{Komissarov}}, \binits{S.S.}},
\bauthor{\bsnm{{Falle}}, \binits{S.A.E.G.}}:
\batitle{{Simulations of Superluminal Radio Sources}}.
\bjtitle{\mnras}
\bvolume{288},
\bfpage{833}--\blpage{848}
(\byear{1997})
\end{barticle}
\endbibitem

\bibitem{BBR-84}
\begin{barticle}
\bauthor{\bsnm{{Begelman}}, \binits{M.C.}},
\bauthor{\bsnm{{Blandford}}, \binits{R.D.}},
\bauthor{\bsnm{{Rees}}, \binits{M.J.}}:
\batitle{{Theory of extragalactic radio sources}}.
\bjtitle{Reviews of Modern Physics}
\bvolume{56},
\bfpage{255}--\blpage{351}
(\byear{1984}).
doi:\doiurl{10.1103/RevModPhys.56.255}
\end{barticle}
\endbibitem

\bibitem{BHK12}
\begin{bbook}
\bauthor{\bsnm{{Boettcher}}, \binits{M.}},
\bauthor{\bsnm{{Harris}}, \binits{D.E.}},
\bauthor{\bsnm{{Krawczynski}}, \binits{H.}}:
\bbtitle{{Relativistic Jets from Active Galactic Nuclei}},
(\byear{2012})
\end{bbook}
\endbibitem

\bibitem{Bz}
\begin{barticle}
\bauthor{\bsnm{{Blandford}}, \binits{R.D.}},
\bauthor{\bsnm{{Znajek}}, \binits{R.L.}}:
\batitle{{Electromagnetic extraction of energy from Kerr black holes}}.
\bjtitle{\mnras}
\bvolume{179},
\bfpage{433}--\blpage{456}
(\byear{1977}).
doi:\doiurl{10.1093/mnras/179.3.433}
\end{barticle}
\endbibitem

\bibitem{MSO-08}
\begin{barticle}
\bauthor{\bsnm{{Moll}}, \binits{R.}},
\bauthor{\bsnm{{Spruit}}, \binits{H.C.}},
\bauthor{\bsnm{{Obergaulinger}}, \binits{M.}}:
\batitle{{Kink instabilities in jets from rotating magnetic fields}}.
\bjtitle{\aap}
\bvolume{492},
\bfpage{621}--\blpage{630}
(\byear{2008}).
doi:\doiurl{10.1051/0004-6361:200810523}.
\arxivurl{0809.3165}
\end{barticle}
\endbibitem

\bibitem{PK-15}
\begin{barticle}
\bauthor{\bsnm{{Porth}}, \binits{O.}},
\bauthor{\bsnm{{Komissarov}}, \binits{S.S.}}:
\batitle{{Causality and stability of cosmic jets}}.
\bjtitle{\mnras}
\bvolume{452},
\bfpage{1089}--\blpage{1104}
(\byear{2015}).
doi:\doiurl{10.1093/mnras/stv1295}.
\arxivurl{1408.3318}
\end{barticle}
\endbibitem

\bibitem{sanders-83}
\begin{barticle}
\bauthor{\bsnm{{Sanders}}, \binits{R.H.}}:
\batitle{{The reconfinement of jets}}.
\bjtitle{\apj}
\bvolume{266},
\bfpage{73}--\blpage{81}
(\byear{1983}).
doi:\doiurl{10.1086/160760}
\end{barticle}
\endbibitem

\bibitem{phinney-83}
\begin{botherref}
\oauthor{\bsnm{{Phinney}}, \binits{E.S.}}
PhD thesis,
, Univ.~Cambridge, (1983)
(1983)
\end{botherref}
\endbibitem

\bibitem{Mathews}
\begin{barticle}
\bauthor{\bsnm{{Mathews}}, \binits{W.G.}},
\bauthor{\bsnm{{Brighenti}}, \binits{F.}}:
\batitle{{Hot Gas in and around Elliptical Galaxies}}.
\bjtitle{\araa}
\bvolume{41},
\bfpage{191}--\blpage{239}
(\byear{2003}).
doi:\doiurl{10.1146/annurev.astro.41.090401.094542}.
\arxivurl{astro-ph/0309553}
\end{barticle}
\endbibitem

\bibitem{BHL-94}
\begin{barticle}
\bauthor{\bsnm{{Bridle}}, \binits{A.H.}},
\bauthor{\bsnm{{Hough}}, \binits{D.H.}},
\bauthor{\bsnm{{Lonsdale}}, \binits{C.J.}},
\bauthor{\bsnm{{Burns}}, \binits{J.O.}},
\bauthor{\bsnm{{Laing}}, \binits{R.A.}}:
\batitle{{Deep VLA imaging of twelve extended 3CR quasars}}.
\bjtitle{\aj}
\bvolume{108},
\bfpage{766}--\blpage{820}
(\byear{1994}).
doi:\doiurl{10.1086/117112}
\end{barticle}
\endbibitem

\bibitem{WF-85}
\begin{barticle}
\bauthor{\bsnm{{Wilson}}, \binits{M.J.}},
\bauthor{\bsnm{{Falle}}, \binits{S.A.E.G.}}:
\batitle{{Steady jets}}.
\bjtitle{\mnras}
\bvolume{216},
\bfpage{971}--\blpage{985}
(\byear{1985})
\end{barticle}
\endbibitem

\bibitem{FW-85}
\begin{barticle}
\bauthor{\bsnm{{Falle}}, \binits{S.A.E.G.}},
\bauthor{\bsnm{{Wilson}}, \binits{M.J.}}:
\batitle{{A theoretical model of the M87 jet}}.
\bjtitle{\mnras}
\bvolume{216},
\bfpage{79}--\blpage{84}
(\byear{1985})
\end{barticle}
\endbibitem

\bibitem{JMM-17}
\begin{barticle}
\bauthor{\bsnm{{Jorstad}}, \binits{S.G.}},
\bauthor{\bsnm{{Marscher}}, \binits{A.P.}},
\bauthor{\bsnm{{Morozova}}, \binits{D.A.}},
\bauthor{\bsnm{{Troitsky}}, \binits{I.S.}},
\bauthor{\bsnm{{Agudo}}, \binits{I.}},
\bauthor{\bsnm{{Casadio}}, \binits{C.}},
\bauthor{\bsnm{{Foord}}, \binits{A.}},
\bauthor{\bsnm{{G{\'o}mez}}, \binits{J.L.}},
\bauthor{\bsnm{{MacDonald}}, \binits{N.R.}},
\bauthor{\bsnm{{Molina}}, \binits{S.N.}},
\bauthor{\bsnm{{L{\"a}hteenm{\"a}ki}}, \binits{A.}},
\bauthor{\bsnm{{Tammi}}, \binits{J.}},
\bauthor{\bsnm{{Tornikoski}}, \binits{M.}}:
\batitle{{Kinematics of Parsec-scale Jets of Gamma-Ray Blazars at 43 GHz within
  the VLBA-BU-BLAZAR Program}}.
\bjtitle{\apj}
\bvolume{846},
\bfpage{98}
(\byear{2017}).
doi:\doiurl{10.3847/1538-4357/aa8407}.
\arxivurl{1711.03983}
\end{barticle}
\endbibitem

\bibitem{wilson-87}
\begin{barticle}
\bauthor{\bsnm{{Wilson}}, \binits{M.J.}}:
\batitle{{Steady relativistic fluid jets}}.
\bjtitle{\mnras}
\bvolume{226},
\bfpage{447}--\blpage{454}
(\byear{1987})
\end{barticle}
\endbibitem

\bibitem{DM-88}
\begin{barticle}
\bauthor{\bsnm{{Daly}}, \binits{R.A.}},
\bauthor{\bsnm{{Marscher}}, \binits{A.P.}}:
\batitle{{The gasdynamics of compact relativistic jets}}.
\bjtitle{\apj}
\bvolume{334},
\bfpage{539}--\blpage{551}
(\byear{1988}).
doi:\doiurl{10.1086/166858}
\end{barticle}
\endbibitem

\bibitem{DP-93}
\begin{barticle}
\bauthor{\bsnm{{Dubal}}, \binits{M.R.}},
\bauthor{\bsnm{{Pantano}}, \binits{O.}}:
\batitle{{The steady-state structure of relativistic magnetic jets}}.
\bjtitle{\mnras}
\bvolume{261},
\bfpage{203}--\blpage{221}
(\byear{1993})
\end{barticle}
\endbibitem

\bibitem{SAK-06}
\begin{barticle}
\bauthor{\bsnm{{Stawarz}}, \binits{{\L}.}},
\bauthor{\bsnm{{Aharonian}}, \binits{F.}},
\bauthor{\bsnm{{Kataoka}}, \binits{J.}},
\bauthor{\bsnm{{Ostrowski}}, \binits{M.}},
\bauthor{\bsnm{{Siemiginowska}}, \binits{A.}},
\bauthor{\bsnm{{Sikora}}, \binits{M.}}:
\batitle{{Dynamics and high-energy emission of the flaring HST-1 knot in the M
  87 jet}}.
\bjtitle{\mnras}
\bvolume{370},
\bfpage{981}--\blpage{992}
(\byear{2006}).
doi:\doiurl{10.1111/j.1365-2966.2006.10525.x}.
\arxivurl{astro-ph/0602220}
\end{barticle}
\endbibitem

\bibitem{NS-09}
\begin{barticle}
\bauthor{\bsnm{{Nalewajko}}, \binits{K.}},
\bauthor{\bsnm{{Sikora}}, \binits{M.}}:
\batitle{{A structure and energy dissipation efficiency of relativistic
  reconfinement shocks}}.
\bjtitle{\mnras}
\bvolume{392},
\bfpage{1205}--\blpage{1210}
(\byear{2009}).
doi:\doiurl{10.1111/j.1365-2966.2008.14123.x}.
\arxivurl{0810.3912}
\end{barticle}
\endbibitem

\bibitem{Nalewajko-12}
\begin{barticle}
\bauthor{\bsnm{{Nalewajko}}, \binits{K.}}:
\batitle{{Dissipation efficiency of reconfinement shocks in relativistic
  jets}}.
\bjtitle{\mnras}
\bvolume{420},
\bfpage{48}--\blpage{52}
(\byear{2012}).
doi:\doiurl{10.1111/j.1745-3933.2011.01193.x}.
\arxivurl{1111.0018}
\end{barticle}
\endbibitem

\bibitem{BL-07}
\begin{barticle}
\bauthor{\bsnm{{Bromberg}}, \binits{O.}},
\bauthor{\bsnm{{Levinson}}, \binits{A.}}:
\batitle{{Hydrodynamic Collimation of Relativistic Outflows: Semianalytic
  Solutions and Application to Gamma-Ray Bursts}}.
\bjtitle{\apj}
\bvolume{671},
\bfpage{678}--\blpage{688}
(\byear{2007}).
doi:\doiurl{10.1086/522668}.
\arxivurl{0705.2040}
\end{barticle}
\endbibitem

\bibitem{BL-09}
\begin{barticle}
\bauthor{\bsnm{{Bromberg}}, \binits{O.}},
\bauthor{\bsnm{{Levinson}}, \binits{A.}}:
\batitle{{Recollimation and Radiative Focusing of Relativistic Jets:
  Applications to Blazars and M87}}.
\bjtitle{\apj}
\bvolume{699},
\bfpage{1274}--\blpage{1280}
(\byear{2009}).
doi:\doiurl{10.1088/0004-637X/699/2/1274}.
\arxivurl{0810.0562}
\end{barticle}
\endbibitem

\bibitem{KB-12}
\begin{barticle}
\bauthor{\bsnm{{Kohler}}, \binits{S.}},
\bauthor{\bsnm{{Begelman}}, \binits{M.C.}},
\bauthor{\bsnm{{Beckwith}}, \binits{K.}}:
\batitle{{Recollimation boundary layers in relativistic jets}}.
\bjtitle{\mnras}
\bvolume{422},
\bfpage{2282}--\blpage{2290}
(\byear{2012}).
doi:\doiurl{10.1111/j.1365-2966.2012.20776.x}.
\arxivurl{1112.4843}
\end{barticle}
\endbibitem

\bibitem{KB-12a}
\begin{barticle}
\bauthor{\bsnm{{Kohler}}, \binits{S.}},
\bauthor{\bsnm{{Begelman}}, \binits{M.C.}}:
\batitle{{Magnetic domination of recollimation boundary layers in relativistic
  jets}}.
\bjtitle{\mnras}
\bvolume{426},
\bfpage{595}--\blpage{600}
(\byear{2012}).
doi:\doiurl{10.1111/j.1365-2966.2012.21876.x}.
\arxivurl{1208.1261}
\end{barticle}
\endbibitem

\bibitem{KB-15}
\begin{barticle}
\bauthor{\bsnm{{Kohler}}, \binits{S.}},
\bauthor{\bsnm{{Begelman}}, \binits{M.C.}}:
\batitle{{Entropy production in relativistic jet boundary layers}}.
\bjtitle{\mnras}
\bvolume{446},
\bfpage{1195}--\blpage{1202}
(\byear{2015}).
doi:\doiurl{10.1093/mnras/stu2135}.
\arxivurl{1411.0666}
\end{barticle}
\endbibitem

\bibitem{MM13}
\begin{barticle}
\bauthor{\bsnm{{Matsumoto}}, \binits{J.}},
\bauthor{\bsnm{{Masada}}, \binits{Y.}}:
\batitle{{Two-dimensional Numerical Study for Rayleigh-Taylor and
  Richtmyer-Meshkov Instabilities in Relativistic Jets}}.
\bjtitle{\apjl}
\bvolume{772},
\bfpage{1}
(\byear{2013}).
doi:\doiurl{10.1088/2041-8205/772/1/L1}.
\arxivurl{1306.1046}
\end{barticle}
\endbibitem

\bibitem{MAP17}
\begin{barticle}
\bauthor{\bsnm{{Matsumoto}}, \binits{J.}},
\bauthor{\bsnm{{Aloy}}, \binits{M.A.}},
\bauthor{\bsnm{{Perucho}}, \binits{M.}}:
\batitle{{Linear theory of the Rayleigh-Taylor instability at a discontinuous
  surface of a relativistic flow}}.
\bjtitle{\mnras}
\bvolume{472},
\bfpage{1421}--\blpage{1431}
(\byear{2017}).
doi:\doiurl{10.1093/mnras/stx2012}.
\arxivurl{1707.04706}
\end{barticle}
\endbibitem

\bibitem{TK-17}
\begin{barticle}
\bauthor{\bsnm{{Toma}}, \binits{K.}},
\bauthor{\bsnm{{Komissarov}}, \binits{S.S.}},
\bauthor{\bsnm{{Porth}}, \binits{O.}}:
\batitle{{Rayleigh-Taylor instability in two-component relativistic jets}}.
\bjtitle{\mnras}
\bvolume{472},
\bfpage{1253}--\blpage{1258}
(\byear{2017}).
doi:\doiurl{10.1093/mnras/stx1770}.
\arxivurl{1705.10425}
\end{barticle}
\endbibitem

\bibitem{GK-18NatAs}
\begin{barticle}
\bauthor{\bsnm{{Gourgouliatos}}, \binits{K.N.}},
\bauthor{\bsnm{{Komissarov}}, \binits{S.S.}}:
\batitle{{Reconfinement and loss of stability in jets from active galactic
  nuclei}}.
\bjtitle{Nature Astronomy}
\bvolume{2},
\bfpage{167}--\blpage{171}
(\byear{2018}).
doi:\doiurl{10.1038/s41550-017-0338-3}
\end{barticle}
\endbibitem

\bibitem{GK-18}
\begin{barticle}
\bauthor{\bsnm{{Gourgouliatos}}, \binits{K.N.}},
\bauthor{\bsnm{{Komissarov}}, \binits{S.S.}}:
\batitle{{Relativistic centrifugal instability}}.
\bjtitle{\mnras}
\bvolume{475},
\bfpage{125}--\blpage{129}
(\byear{2018}).
doi:\doiurl{10.1093/mnrasl/sly016}.
\arxivurl{1710.01345}
\end{barticle}
\endbibitem

\bibitem{FR-74}
\begin{barticle}
\bauthor{\bsnm{{Fanaroff}}, \binits{B.L.}},
\bauthor{\bsnm{{Riley}}, \binits{J.M.}}:
\batitle{{The morphology of extragalactic radio sources of high and low
  luminosity}}.
\bjtitle{\mnras}
\bvolume{167},
\bfpage{31}--\blpage{36}
(\byear{1974})
\end{barticle}
\endbibitem

\bibitem{F-91}
\begin{barticle}
\bauthor{\bsnm{{Falle}}, \binits{S.A.E.G.}}:
\batitle{{Self-similar jets}}.
\bjtitle{\mnras}
\bvolume{250},
\bfpage{581}--\blpage{596}
(\byear{1991})
\end{barticle}
\endbibitem

\bibitem{Komp-60}
\begin{barticle}
\bauthor{\bsnm{{Kompaneets}}, \binits{A.S.}}:
\batitle{{A Point Explosion in an Inhomogeneous Atmosphere}}.
\bjtitle{Soviet Physics Doklady}
\bvolume{5},
\bfpage{46}
(\byear{1960})
\end{barticle}
\endbibitem

\bibitem{KPL-15}
\begin{barticle}
\bauthor{\bsnm{{Komissarov}}, \binits{S.S.}},
\bauthor{\bsnm{{Porth}}, \binits{O.}},
\bauthor{\bsnm{{Lyutikov}}, \binits{M.}}:
\batitle{{Stationary relativistic jets}}.
\bjtitle{Computational Astrophysics and Cosmology}
\bvolume{2},
\bfpage{9}
(\byear{2015}).
doi:\doiurl{10.1186/s40668-015-0013-y}.
\arxivurl{1504.07534}
\end{barticle}
\endbibitem

\bibitem{UKRCL-99}
\begin{barticle}
\bauthor{\bsnm{{Ustyugova}}, \binits{G.V.}},
\bauthor{\bsnm{{Koldoba}}, \binits{A.V.}},
\bauthor{\bsnm{{Romanova}}, \binits{M.M.}},
\bauthor{\bsnm{{Chechetkin}}, \binits{V.M.}},
\bauthor{\bsnm{{Lovelace}}, \binits{R.V.E.}}:
\batitle{{Magnetocentrifugally Driven Winds: Comparison of MHD Simulations with
  Theory}}.
\bjtitle{\apj}
\bvolume{516},
\bfpage{221}--\blpage{235}
(\byear{1999}).
doi:\doiurl{10.1086/307093}.
\arxivurl{astro-ph/9812284}
\end{barticle}
\endbibitem

\bibitem{kvkb-09}
\begin{barticle}
\bauthor{\bsnm{{Komissarov}}, \binits{S.S.}},
\bauthor{\bsnm{{Vlahakis}}, \binits{N.}},
\bauthor{\bsnm{{K{\"o}nigl}}, \binits{A.}},
\bauthor{\bsnm{{Barkov}}, \binits{M.V.}}:
\batitle{{Magnetic acceleration of ultrarelativistic jets in gamma-ray burst
  sources}}.
\bjtitle{\mnras}
\bvolume{394},
\bfpage{1182}--\blpage{1212}
(\byear{2009}).
doi:\doiurl{10.1111/j.1365-2966.2009.14410.x}.
\arxivurl{0811.1467}
\end{barticle}
\endbibitem

\bibitem{TMN-08}
\begin{barticle}
\bauthor{\bsnm{{Tchekhovskoy}}, \binits{A.}},
\bauthor{\bsnm{{McKinney}}, \binits{J.C.}},
\bauthor{\bsnm{{Narayan}}, \binits{R.}}:
\batitle{{Simulations of ultrarelativistic magnetodynamic jets from gamma-ray
  burst engines}}.
\bjtitle{\mnras}
\bvolume{388},
\bfpage{551}--\blpage{572}
(\byear{2008}).
doi:\doiurl{10.1111/j.1365-2966.2008.13425.x}.
\arxivurl{0803.3807}
\end{barticle}
\endbibitem

\bibitem{ssk-godun99}
\begin{barticle}
\bauthor{\bsnm{{Komissarov}}, \binits{S.S.}}:
\batitle{{A Godunov-type scheme for relativistic magnetohydrodynamics}}.
\bjtitle{\mnras}
\bvolume{303},
\bfpage{343}--\blpage{366}
(\byear{1999}).
doi:\doiurl{10.1046/j.1365-8711.1999.02244.x}
\end{barticle}
\endbibitem

\bibitem{King1972}
\begin{barticle}
\bauthor{\bsnm{{King}}, \binits{I.R.}}:
\batitle{{Density Data and Emission Measure for a Model of the Coma Cluster}}.
\bjtitle{\apjl}
\bvolume{174},
\bfpage{123}
(\byear{1972}).
doi:\doiurl{10.1086/180963}
\end{barticle}
\endbibitem

\bibitem{Wang-13}
\begin{barticle}
\bauthor{\bsnm{{Wang}}, \binits{Q.D.}},
\bauthor{\bsnm{{Nowak}}, \binits{M.A.}},
\bauthor{\bsnm{{Markoff}}, \binits{S.B.}},
\bauthor{\bsnm{{Baganoff}}, \binits{F.K.}},
\bauthor{\bsnm{{Nayakshin}}, \binits{S.}},
\bauthor{\bsnm{{Yuan}}, \binits{F.}},
\bauthor{\bsnm{{Cuadra}}, \binits{J.}},
\bauthor{\bsnm{{Davis}}, \binits{J.}},
\bauthor{\bsnm{{Dexter}}, \binits{J.}},
\bauthor{\bsnm{{Fabian}}, \binits{A.C.}},
\bauthor{\bsnm{{Grosso}}, \binits{N.}},
\bauthor{\bsnm{{Haggard}}, \binits{D.}},
\bauthor{\bsnm{{Houck}}, \binits{J.}},
\bauthor{\bsnm{{Ji}}, \binits{L.}},
\bauthor{\bsnm{{Li}}, \binits{Z.}},
\bauthor{\bsnm{{Neilsen}}, \binits{J.}},
\bauthor{\bsnm{{Porquet}}, \binits{D.}},
\bauthor{\bsnm{{Ripple}}, \binits{F.}},
\bauthor{\bsnm{{Shcherbakov}}, \binits{R.V.}}:
\batitle{{Dissecting X-ray-Emitting Gas Around the Center of Our Galaxy}}.
\bjtitle{Science}
\bvolume{341},
\bfpage{981}--\blpage{983}
(\byear{2013}).
doi:\doiurl{10.1126/science.1240755}.
\arxivurl{1307.5845}
\end{barticle}
\endbibitem

\bibitem{Wong-14}
\begin{barticle}
\bauthor{\bsnm{{Wong}}, \binits{K.-W.}},
\bauthor{\bsnm{{Irwin}}, \binits{J.A.}},
\bauthor{\bsnm{{Shcherbakov}}, \binits{R.V.}},
\bauthor{\bsnm{{Yukita}}, \binits{M.}},
\bauthor{\bsnm{{Million}}, \binits{E.T.}},
\bauthor{\bsnm{{Bregman}}, \binits{J.N.}}:
\batitle{{The Megasecond Chandra X-Ray Visionary Project Observation of NGC
  3115: Witnessing the Flow of Hot Gas within the Bondi Radius}}.
\bjtitle{\apj}
\bvolume{780},
\bfpage{9}
(\byear{2014}).
doi:\doiurl{10.1088/0004-637X/780/1/9}.
\arxivurl{1311.0868}
\end{barticle}
\endbibitem

\bibitem{Russell-15}
\begin{barticle}
\bauthor{\bsnm{{Russell}}, \binits{H.R.}},
\bauthor{\bsnm{{Fabian}}, \binits{A.C.}},
\bauthor{\bsnm{{McNamara}}, \binits{B.R.}},
\bauthor{\bsnm{{Broderick}}, \binits{A.E.}}:
\batitle{{Inside the Bondi radius of M87}}.
\bjtitle{\mnras}
\bvolume{451},
\bfpage{588}--\blpage{600}
(\byear{2015}).
doi:\doiurl{10.1093/mnras/stv954}.
\arxivurl{1504.07633}
\end{barticle}
\endbibitem

\bibitem{RPFF90}
\begin{barticle}
\bauthor{\bsnm{{de Ruiter}}, \binits{H.R.}},
\bauthor{\bsnm{{Parma}}, \binits{P.}},
\bauthor{\bsnm{{Fanti}}, \binits{C.}},
\bauthor{\bsnm{{Fanti}}, \binits{R.}}:
\batitle{{VLA observations of low luminosity radio galaxies. VII - General
  properties}}.
\bjtitle{\aap}
\bvolume{227},
\bfpage{351}--\blpage{361}
(\byear{1990})
\end{barticle}
\endbibitem

\bibitem{Cavagnolo}
\begin{barticle}
\bauthor{\bsnm{{Cavagnolo}}, \binits{K.W.}},
\bauthor{\bsnm{{McNamara}}, \binits{B.R.}},
\bauthor{\bsnm{{Nulsen}}, \binits{P.E.J.}},
\bauthor{\bsnm{{Carilli}}, \binits{C.L.}},
\bauthor{\bsnm{{Jones}}, \binits{C.}},
\bauthor{\bsnm{{B{\^i}rzan}}, \binits{L.}}:
\batitle{{A Relationship Between AGN Jet Power and Radio Power}}.
\bjtitle{\apj}
\bvolume{720},
\bfpage{1066}--\blpage{1072}
(\byear{2010}).
doi:\doiurl{10.1088/0004-637X/720/2/1066}.
\arxivurl{1006.5699}
\end{barticle}
\endbibitem

\bibitem{OL-94}
\begin{bchapter}
\bauthor{\bsnm{{Owen}}, \binits{F.N.}},
\bauthor{\bsnm{{Ledlow}}, \binits{M.J.}}:
\bctitle{{The FRI/Il Break and the Bivariate Luminosity Function in Abell
  Clusters of Galaxies}}.
In: \beditor{\bsnm{{Bicknell}}, \binits{G.V.}},
\beditor{\bsnm{{Dopita}}, \binits{M.A.}},
\beditor{\bsnm{{Quinn}}, \binits{P.J.}} (eds.)
\bbtitle{The Physics of Active Galaxies}.
\bsertitle{Astronomical Society of the Pacific Conference Series},
vol. \bseriesno{54},
p. \bfpage{319}
(\byear{1994})
\end{bchapter}
\endbibitem

\bibitem{FF90}
\begin{barticle}
\bauthor{\bsnm{{Fanti}}, \binits{R.}},
\bauthor{\bsnm{{Fanti}}, \binits{C.}},
\bauthor{\bsnm{{Schilizzi}}, \binits{R.T.}},
\bauthor{\bsnm{{Spencer}}, \binits{R.E.}},
\bauthor{\bsnm{{Nan Rendong}}},
\bauthor{\bsnm{{Parma}}, \binits{P.}},
\bauthor{\bsnm{{van Breugel}}, \binits{W.J.M.}},
\bauthor{\bsnm{{Venturi}}, \binits{T.}}:
\batitle{{On the nature of compact steep spectrum radio sources.}}
\bjtitle{\aap}
\bvolume{231},
\bfpage{333}--\blpage{346}
(\byear{1990})
\end{barticle}
\endbibitem

\bibitem{MBR16}
\begin{barticle}
\bauthor{\bsnm{{Massaglia}}, \binits{S.}},
\bauthor{\bsnm{{Bodo}}, \binits{G.}},
\bauthor{\bsnm{{Rossi}}, \binits{P.}},
\bauthor{\bsnm{{Capetti}}, \binits{S.}},
\bauthor{\bsnm{{Mignone}}, \binits{A.}}:
\batitle{{Making Faranoff-Riley I radio sources. I. Numerical hydrodynamic 3D
  simulations of low-power jets}}.
\bjtitle{\aap}
\bvolume{596},
\bfpage{12}
(\byear{2016}).
doi:\doiurl{10.1051/0004-6361/201629375}.
\arxivurl{1609.02497}
\end{barticle}
\endbibitem

\bibitem{TB16}
\begin{barticle}
\bauthor{\bsnm{{Tchekhovskoy}}, \binits{A.}},
\bauthor{\bsnm{{Bromberg}}, \binits{O.}}:
\batitle{{Three-dimensional relativistic MHD simulations of active galactic
  nuclei jets: magnetic kink instability and Fanaroff-Riley dichotomy}}.
\bjtitle{\mnras}
\bvolume{461},
\bfpage{46}--\blpage{50}
(\byear{2016}).
doi:\doiurl{10.1093/mnrasl/slw064}.
\arxivurl{1512.04526}
\end{barticle}
\endbibitem

\end{thebibliography}

\newcommand{\BMCxmlcomment}[1]{}

\BMCxmlcomment{

<refgrp>

<bibl id="B1">
  <title><p>{Simulations of Superluminal Radio Sources}</p></title>
  <aug>
    <au><snm>{Komissarov}</snm><fnm>S. S.</fnm></au>
    <au><snm>{Falle}</snm><fnm>S. A. E. G.</fnm></au>
  </aug>
  <source>\mnras</source>
  <pubdate>1997</pubdate>
  <volume>288</volume>
  <fpage>833</fpage>
  <lpage>848</lpage>
</bibl>

<bibl id="B2">
  <title><p>{Theory of extragalactic radio sources}</p></title>
  <aug>
    <au><snm>{Begelman}</snm><fnm>M. C.</fnm></au>
    <au><snm>{Blandford}</snm><fnm>R. D.</fnm></au>
    <au><snm>{Rees}</snm><fnm>M. J.</fnm></au>
  </aug>
  <source>Reviews of Modern Physics</source>
  <pubdate>1984</pubdate>
  <volume>56</volume>
  <fpage>255</fpage>
  <lpage>351</lpage>
</bibl>

<bibl id="B3">
  <title><p>{Relativistic Jets from Active Galactic Nuclei}</p></title>
  <aug>
    <au><snm>{Boettcher}</snm><fnm>M.</fnm></au>
    <au><snm>{Harris}</snm><fnm>D. E.</fnm></au>
    <au><snm>{Krawczynski}</snm><fnm>H.</fnm></au>
  </aug>
  <source>Relativistic Jets from Active Galactic Nuclei, by M.~Boettcher,
  D.E.~Harris, ahd H.~Krawczynski, 425 pages.~ Berlin: Wiley, 2012</source>
  <pubdate>2012</pubdate>
</bibl>

<bibl id="B4">
  <title><p>{Electromagnetic extraction of energy from Kerr black
  holes}</p></title>
  <aug>
    <au><snm>{Blandford}</snm><fnm>R. D.</fnm></au>
    <au><snm>{Znajek}</snm><fnm>R. L.</fnm></au>
  </aug>
  <source>\mnras</source>
  <pubdate>1977</pubdate>
  <volume>179</volume>
  <fpage>433</fpage>
  <lpage>456</lpage>
</bibl>

<bibl id="B5">
  <title><p>{Kink instabilities in jets from rotating magnetic
  fields}</p></title>
  <aug>
    <au><snm>{Moll}</snm><fnm>R.</fnm></au>
    <au><snm>{Spruit}</snm><fnm>H. C.</fnm></au>
    <au><snm>{Obergaulinger}</snm><fnm>M.</fnm></au>
  </aug>
  <source>\aap</source>
  <pubdate>2008</pubdate>
  <volume>492</volume>
  <fpage>621</fpage>
  <lpage>630</lpage>
</bibl>

<bibl id="B6">
  <title><p>{Causality and stability of cosmic jets}</p></title>
  <aug>
    <au><snm>{Porth}</snm><fnm>O.</fnm></au>
    <au><snm>{Komissarov}</snm><fnm>S. S.</fnm></au>
  </aug>
  <source>\mnras</source>
  <pubdate>2015</pubdate>
  <volume>452</volume>
  <fpage>1089</fpage>
  <lpage>1104</lpage>
</bibl>

<bibl id="B7">
  <title><p>{The reconfinement of jets}</p></title>
  <aug>
    <au><snm>{Sanders}</snm><fnm>R. H.</fnm></au>
  </aug>
  <source>\apj</source>
  <pubdate>1983</pubdate>
  <volume>266</volume>
  <fpage>73</fpage>
  <lpage>81</lpage>
</bibl>

<bibl id="B8">
  <aug>
    <au><snm>{Phinney}</snm><fnm>E. S.</fnm></au>
  </aug>
  <source>PhD thesis</source>
  <publisher>, Univ.~Cambridge, (1983)</publisher>
  <pubdate>1983</pubdate>
</bibl>

<bibl id="B9">
  <title><p>{Hot Gas in and around Elliptical Galaxies}</p></title>
  <aug>
    <au><snm>{Mathews}</snm><fnm>W. G.</fnm></au>
    <au><snm>{Brighenti}</snm><fnm>F.</fnm></au>
  </aug>
  <source>\araa</source>
  <pubdate>2003</pubdate>
  <volume>41</volume>
  <fpage>191</fpage>
  <lpage>239</lpage>
</bibl>

<bibl id="B10">
  <title><p>{Deep VLA imaging of twelve extended 3CR quasars}</p></title>
  <aug>
    <au><snm>{Bridle}</snm><fnm>A. H.</fnm></au>
    <au><snm>{Hough}</snm><fnm>D. H.</fnm></au>
    <au><snm>{Lonsdale}</snm><fnm>C. J.</fnm></au>
    <au><snm>{Burns}</snm><fnm>J. O.</fnm></au>
    <au><snm>{Laing}</snm><fnm>R. A.</fnm></au>
  </aug>
  <source>\aj</source>
  <pubdate>1994</pubdate>
  <volume>108</volume>
  <fpage>766</fpage>
  <lpage>820</lpage>
</bibl>

<bibl id="B11">
  <title><p>{Steady jets}</p></title>
  <aug>
    <au><snm>{Wilson}</snm><fnm>M. J.</fnm></au>
    <au><snm>{Falle}</snm><fnm>S. A. E. G.</fnm></au>
  </aug>
  <source>\mnras</source>
  <pubdate>1985</pubdate>
  <volume>216</volume>
  <fpage>971</fpage>
  <lpage>985</lpage>
</bibl>

<bibl id="B12">
  <title><p>{A theoretical model of the M87 jet}</p></title>
  <aug>
    <au><snm>{Falle}</snm><fnm>S. A. E. G.</fnm></au>
    <au><snm>{Wilson}</snm><fnm>M. J.</fnm></au>
  </aug>
  <source>\mnras</source>
  <pubdate>1985</pubdate>
  <volume>216</volume>
  <fpage>79</fpage>
  <lpage>84</lpage>
</bibl>

<bibl id="B13">
  <title><p>{Kinematics of Parsec-scale Jets of Gamma-Ray Blazars at 43 GHz
  within the VLBA-BU-BLAZAR Program}</p></title>
  <aug>
    <au><snm>{Jorstad}</snm><fnm>S. G.</fnm></au>
    <au><snm>{Marscher}</snm><fnm>A. P.</fnm></au>
    <au><snm>{Morozova}</snm><fnm>D. A.</fnm></au>
    <au><snm>{Troitsky}</snm><fnm>I. S.</fnm></au>
    <au><snm>{Agudo}</snm><fnm>I.</fnm></au>
    <au><snm>{Casadio}</snm><fnm>C.</fnm></au>
    <au><snm>{Foord}</snm><fnm>A.</fnm></au>
    <au><snm>{G{\'o}mez}</snm><fnm>J. L.</fnm></au>
    <au><snm>{MacDonald}</snm><fnm>N. R.</fnm></au>
    <au><snm>{Molina}</snm><fnm>S. N.</fnm></au>
    <au><snm>{L{\"a}hteenm{\"a}ki}</snm><fnm>A.</fnm></au>
    <au><snm>{Tammi}</snm><fnm>J.</fnm></au>
    <au><snm>{Tornikoski}</snm><fnm>M.</fnm></au>
  </aug>
  <source>\apj</source>
  <pubdate>2017</pubdate>
  <volume>846</volume>
  <fpage>98</fpage>
</bibl>

<bibl id="B14">
  <title><p>{Steady relativistic fluid jets}</p></title>
  <aug>
    <au><snm>{Wilson}</snm><fnm>M. J.</fnm></au>
  </aug>
  <source>\mnras</source>
  <pubdate>1987</pubdate>
  <volume>226</volume>
  <fpage>447</fpage>
  <lpage>454</lpage>
</bibl>

<bibl id="B15">
  <title><p>{The gasdynamics of compact relativistic jets}</p></title>
  <aug>
    <au><snm>{Daly}</snm><fnm>R. A.</fnm></au>
    <au><snm>{Marscher}</snm><fnm>A. P.</fnm></au>
  </aug>
  <source>\apj</source>
  <pubdate>1988</pubdate>
  <volume>334</volume>
  <fpage>539</fpage>
  <lpage>551</lpage>
</bibl>

<bibl id="B16">
  <title><p>{The steady-state structure of relativistic magnetic
  jets}</p></title>
  <aug>
    <au><snm>{Dubal}</snm><fnm>M. R.</fnm></au>
    <au><snm>{Pantano}</snm><fnm>O.</fnm></au>
  </aug>
  <source>\mnras</source>
  <pubdate>1993</pubdate>
  <volume>261</volume>
  <fpage>203</fpage>
  <lpage>221</lpage>
</bibl>

<bibl id="B17">
  <title><p>{Dynamics and high-energy emission of the flaring HST-1 knot in the
  M 87 jet}</p></title>
  <aug>
    <au><snm>{Stawarz}</snm><fnm>{\L}.</fnm></au>
    <au><snm>{Aharonian}</snm><fnm>F.</fnm></au>
    <au><snm>{Kataoka}</snm><fnm>J.</fnm></au>
    <au><snm>{Ostrowski}</snm><fnm>M.</fnm></au>
    <au><snm>{Siemiginowska}</snm><fnm>A.</fnm></au>
    <au><snm>{Sikora}</snm><fnm>M.</fnm></au>
  </aug>
  <source>\mnras</source>
  <pubdate>2006</pubdate>
  <volume>370</volume>
  <fpage>981</fpage>
  <lpage>992</lpage>
</bibl>

<bibl id="B18">
  <title><p>{A structure and energy dissipation efficiency of relativistic
  reconfinement shocks}</p></title>
  <aug>
    <au><snm>{Nalewajko}</snm><fnm>K.</fnm></au>
    <au><snm>{Sikora}</snm><fnm>M.</fnm></au>
  </aug>
  <source>\mnras</source>
  <pubdate>2009</pubdate>
  <volume>392</volume>
  <fpage>1205</fpage>
  <lpage>1210</lpage>
</bibl>

<bibl id="B19">
  <title><p>{Dissipation efficiency of reconfinement shocks in relativistic
  jets}</p></title>
  <aug>
    <au><snm>{Nalewajko}</snm><fnm>K.</fnm></au>
  </aug>
  <source>\mnras</source>
  <pubdate>2012</pubdate>
  <volume>420</volume>
  <fpage>L48</fpage>
  <lpage>L52</lpage>
</bibl>

<bibl id="B20">
  <title><p>{Hydrodynamic Collimation of Relativistic Outflows: Semianalytic
  Solutions and Application to Gamma-Ray Bursts}</p></title>
  <aug>
    <au><snm>{Bromberg}</snm><fnm>O.</fnm></au>
    <au><snm>{Levinson}</snm><fnm>A.</fnm></au>
  </aug>
  <source>\apj</source>
  <pubdate>2007</pubdate>
  <volume>671</volume>
  <fpage>678</fpage>
  <lpage>688</lpage>
</bibl>

<bibl id="B21">
  <title><p>{Recollimation and Radiative Focusing of Relativistic Jets:
  Applications to Blazars and M87}</p></title>
  <aug>
    <au><snm>{Bromberg}</snm><fnm>O.</fnm></au>
    <au><snm>{Levinson}</snm><fnm>A.</fnm></au>
  </aug>
  <source>\apj</source>
  <pubdate>2009</pubdate>
  <volume>699</volume>
  <fpage>1274</fpage>
  <lpage>1280</lpage>
</bibl>

<bibl id="B22">
  <title><p>{Recollimation boundary layers in relativistic jets}</p></title>
  <aug>
    <au><snm>{Kohler}</snm><fnm>S.</fnm></au>
    <au><snm>{Begelman}</snm><fnm>M. C.</fnm></au>
    <au><snm>{Beckwith}</snm><fnm>K.</fnm></au>
  </aug>
  <source>\mnras</source>
  <pubdate>2012</pubdate>
  <volume>422</volume>
  <fpage>2282</fpage>
  <lpage>2290</lpage>
</bibl>

<bibl id="B23">
  <title><p>{Magnetic domination of recollimation boundary layers in
  relativistic jets}</p></title>
  <aug>
    <au><snm>{Kohler}</snm><fnm>S.</fnm></au>
    <au><snm>{Begelman}</snm><fnm>M. C.</fnm></au>
  </aug>
  <source>\mnras</source>
  <pubdate>2012</pubdate>
  <volume>426</volume>
  <fpage>595</fpage>
  <lpage>600</lpage>
</bibl>

<bibl id="B24">
  <title><p>{Entropy production in relativistic jet boundary
  layers}</p></title>
  <aug>
    <au><snm>{Kohler}</snm><fnm>S.</fnm></au>
    <au><snm>{Begelman}</snm><fnm>M. C.</fnm></au>
  </aug>
  <source>\mnras</source>
  <pubdate>2015</pubdate>
  <volume>446</volume>
  <fpage>1195</fpage>
  <lpage>1202</lpage>
</bibl>

<bibl id="B25">
  <title><p>{Two-dimensional Numerical Study for Rayleigh-Taylor and
  Richtmyer-Meshkov Instabilities in Relativistic Jets}</p></title>
  <aug>
    <au><snm>{Matsumoto}</snm><fnm>J.</fnm></au>
    <au><snm>{Masada}</snm><fnm>Y.</fnm></au>
  </aug>
  <source>\apjl</source>
  <pubdate>2013</pubdate>
  <volume>772</volume>
  <fpage>L1</fpage>
</bibl>

<bibl id="B26">
  <title><p>{Linear theory of the Rayleigh-Taylor instability at a
  discontinuous surface of a relativistic flow}</p></title>
  <aug>
    <au><snm>{Matsumoto}</snm><fnm>J.</fnm></au>
    <au><snm>{Aloy}</snm><fnm>M. A.</fnm></au>
    <au><snm>{Perucho}</snm><fnm>M.</fnm></au>
  </aug>
  <source>\mnras</source>
  <pubdate>2017</pubdate>
  <volume>472</volume>
  <fpage>1421</fpage>
  <lpage>1431</lpage>
</bibl>

<bibl id="B27">
  <title><p>{Rayleigh-Taylor instability in two-component relativistic
  jets}</p></title>
  <aug>
    <au><snm>{Toma}</snm><fnm>K.</fnm></au>
    <au><snm>{Komissarov}</snm><fnm>S. S.</fnm></au>
    <au><snm>{Porth}</snm><fnm>O.</fnm></au>
  </aug>
  <source>\mnras</source>
  <pubdate>2017</pubdate>
  <volume>472</volume>
  <fpage>1253</fpage>
  <lpage>1258</lpage>
</bibl>

<bibl id="B28">
  <title><p>{Reconfinement and loss of stability in jets from active galactic
  nuclei}</p></title>
  <aug>
    <au><snm>{Gourgouliatos}</snm><fnm>K. N.</fnm></au>
    <au><snm>{Komissarov}</snm><fnm>S. S.</fnm></au>
  </aug>
  <source>Nature Astronomy</source>
  <pubdate>2018</pubdate>
  <volume>2</volume>
  <fpage>167</fpage>
  <lpage>171</lpage>
</bibl>

<bibl id="B29">
  <title><p>{Relativistic centrifugal instability}</p></title>
  <aug>
    <au><snm>{Gourgouliatos}</snm><fnm>K. N.</fnm></au>
    <au><snm>{Komissarov}</snm><fnm>S. S.</fnm></au>
  </aug>
  <source>\mnras</source>
  <pubdate>2018</pubdate>
  <volume>475</volume>
  <fpage>L125</fpage>
  <lpage>L129</lpage>
</bibl>

<bibl id="B30">
  <title><p>{The morphology of extragalactic radio sources of high and low
  luminosity}</p></title>
  <aug>
    <au><snm>{Fanaroff}</snm><fnm>B. L.</fnm></au>
    <au><snm>{Riley}</snm><fnm>J. M.</fnm></au>
  </aug>
  <source>\mnras</source>
  <pubdate>1974</pubdate>
  <volume>167</volume>
  <fpage>31P</fpage>
  <lpage>36P</lpage>
</bibl>

<bibl id="B31">
  <title><p>{Self-similar jets}</p></title>
  <aug>
    <au><snm>{Falle}</snm><fnm>S. A. E. G.</fnm></au>
  </aug>
  <source>\mnras</source>
  <pubdate>1991</pubdate>
  <volume>250</volume>
  <fpage>581</fpage>
  <lpage>596</lpage>
</bibl>

<bibl id="B32">
  <title><p>{A Point Explosion in an Inhomogeneous Atmosphere}</p></title>
  <aug>
    <au><snm>{Kompaneets}</snm><fnm>A. S.</fnm></au>
  </aug>
  <source>Soviet Physics Doklady</source>
  <pubdate>1960</pubdate>
  <volume>5</volume>
  <fpage>46</fpage>
</bibl>

<bibl id="B33">
  <title><p>{Stationary relativistic jets}</p></title>
  <aug>
    <au><snm>{Komissarov}</snm><fnm>S. S.</fnm></au>
    <au><snm>{Porth}</snm><fnm>O.</fnm></au>
    <au><snm>{Lyutikov}</snm><fnm>M.</fnm></au>
  </aug>
  <source>Computational Astrophysics and Cosmology</source>
  <pubdate>2015</pubdate>
  <volume>2</volume>
  <fpage>9</fpage>
</bibl>

<bibl id="B34">
  <title><p>{Magnetocentrifugally Driven Winds: Comparison of MHD Simulations
  with Theory}</p></title>
  <aug>
    <au><snm>{Ustyugova}</snm><fnm>G. V.</fnm></au>
    <au><snm>{Koldoba}</snm><fnm>A. V.</fnm></au>
    <au><snm>{Romanova}</snm><fnm>M. M.</fnm></au>
    <au><snm>{Chechetkin}</snm><fnm>V. M.</fnm></au>
    <au><snm>{Lovelace}</snm><fnm>R. V. E.</fnm></au>
  </aug>
  <source>\apj</source>
  <pubdate>1999</pubdate>
  <volume>516</volume>
  <fpage>221</fpage>
  <lpage>235</lpage>
</bibl>

<bibl id="B35">
  <title><p>{Magnetic acceleration of ultrarelativistic jets in gamma-ray burst
  sources}</p></title>
  <aug>
    <au><snm>{Komissarov}</snm><fnm>S. S.</fnm></au>
    <au><snm>{Vlahakis}</snm><fnm>N.</fnm></au>
    <au><snm>{K{\"o}nigl}</snm><fnm>A.</fnm></au>
    <au><snm>{Barkov}</snm><fnm>M. V.</fnm></au>
  </aug>
  <source>\mnras</source>
  <pubdate>2009</pubdate>
  <volume>394</volume>
  <fpage>1182</fpage>
  <lpage>1212</lpage>
</bibl>

<bibl id="B36">
  <title><p>{Simulations of ultrarelativistic magnetodynamic jets from
  gamma-ray burst engines}</p></title>
  <aug>
    <au><snm>{Tchekhovskoy}</snm><fnm>A.</fnm></au>
    <au><snm>{McKinney}</snm><fnm>J. C.</fnm></au>
    <au><snm>{Narayan}</snm><fnm>R.</fnm></au>
  </aug>
  <source>\mnras</source>
  <pubdate>2008</pubdate>
  <volume>388</volume>
  <fpage>551</fpage>
  <lpage>572</lpage>
</bibl>

<bibl id="B37">
  <title><p>{A Godunov-type scheme for relativistic
  magnetohydrodynamics}</p></title>
  <aug>
    <au><snm>{Komissarov}</snm><fnm>S. S.</fnm></au>
  </aug>
  <source>\mnras</source>
  <pubdate>1999</pubdate>
  <volume>303</volume>
  <fpage>343</fpage>
  <lpage>366</lpage>
</bibl>

<bibl id="B38">
  <title><p>{Density Data and Emission Measure for a Model of the Coma
  Cluster}</p></title>
  <aug>
    <au><snm>{King}</snm><fnm>I. R.</fnm></au>
  </aug>
  <source>\apjl</source>
  <pubdate>1972</pubdate>
  <volume>174</volume>
  <fpage>L123</fpage>
</bibl>

<bibl id="B39">
  <title><p>{Dissecting X-ray-Emitting Gas Around the Center of Our
  Galaxy}</p></title>
  <aug>
    <au><snm>{Wang}</snm><fnm>Q. D.</fnm></au>
    <au><snm>{Nowak}</snm><fnm>M. A.</fnm></au>
    <au><snm>{Markoff}</snm><fnm>S. B.</fnm></au>
    <au><snm>{Baganoff}</snm><fnm>F. K.</fnm></au>
    <au><snm>{Nayakshin}</snm><fnm>S.</fnm></au>
    <au><snm>{Yuan}</snm><fnm>F.</fnm></au>
    <au><snm>{Cuadra}</snm><fnm>J.</fnm></au>
    <au><snm>{Davis}</snm><fnm>J.</fnm></au>
    <au><snm>{Dexter}</snm><fnm>J.</fnm></au>
    <au><snm>{Fabian}</snm><fnm>A. C.</fnm></au>
    <au><snm>{Grosso}</snm><fnm>N.</fnm></au>
    <au><snm>{Haggard}</snm><fnm>D.</fnm></au>
    <au><snm>{Houck}</snm><fnm>J.</fnm></au>
    <au><snm>{Ji}</snm><fnm>L.</fnm></au>
    <au><snm>{Li}</snm><fnm>Z.</fnm></au>
    <au><snm>{Neilsen}</snm><fnm>J.</fnm></au>
    <au><snm>{Porquet}</snm><fnm>D.</fnm></au>
    <au><snm>{Ripple}</snm><fnm>F.</fnm></au>
    <au><snm>{Shcherbakov}</snm><fnm>R. V.</fnm></au>
  </aug>
  <source>Science</source>
  <pubdate>2013</pubdate>
  <volume>341</volume>
  <fpage>981</fpage>
  <lpage>983</lpage>
</bibl>

<bibl id="B40">
  <title><p>{The Megasecond Chandra X-Ray Visionary Project Observation of NGC
  3115: Witnessing the Flow of Hot Gas within the Bondi Radius}</p></title>
  <aug>
    <au><snm>{Wong}</snm><fnm>K. W.</fnm></au>
    <au><snm>{Irwin}</snm><fnm>J. A.</fnm></au>
    <au><snm>{Shcherbakov}</snm><fnm>R. V.</fnm></au>
    <au><snm>{Yukita}</snm><fnm>M.</fnm></au>
    <au><snm>{Million}</snm><fnm>E. T.</fnm></au>
    <au><snm>{Bregman}</snm><fnm>J. N.</fnm></au>
  </aug>
  <source>\apj</source>
  <pubdate>2014</pubdate>
  <volume>780</volume>
  <fpage>9</fpage>
</bibl>

<bibl id="B41">
  <title><p>{Inside the Bondi radius of M87}</p></title>
  <aug>
    <au><snm>{Russell}</snm><fnm>H. R.</fnm></au>
    <au><snm>{Fabian}</snm><fnm>A. C.</fnm></au>
    <au><snm>{McNamara}</snm><fnm>B. R.</fnm></au>
    <au><snm>{Broderick}</snm><fnm>A. E.</fnm></au>
  </aug>
  <source>\mnras</source>
  <pubdate>2015</pubdate>
  <volume>451</volume>
  <fpage>588</fpage>
  <lpage>600</lpage>
</bibl>

<bibl id="B42">
  <title><p>{VLA observations of low luminosity radio galaxies. VII - General
  properties}</p></title>
  <aug>
    <au><snm>{de Ruiter}</snm><fnm>H. R.</fnm></au>
    <au><snm>{Parma}</snm><fnm>P.</fnm></au>
    <au><snm>{Fanti}</snm><fnm>C.</fnm></au>
    <au><snm>{Fanti}</snm><fnm>R.</fnm></au>
  </aug>
  <source>\aap</source>
  <pubdate>1990</pubdate>
  <volume>227</volume>
  <fpage>351</fpage>
  <lpage>361</lpage>
</bibl>

<bibl id="B43">
  <title><p>{A Relationship Between AGN Jet Power and Radio Power}</p></title>
  <aug>
    <au><snm>{Cavagnolo}</snm><fnm>K. W.</fnm></au>
    <au><snm>{McNamara}</snm><fnm>B. R.</fnm></au>
    <au><snm>{Nulsen}</snm><fnm>P. E. J.</fnm></au>
    <au><snm>{Carilli}</snm><fnm>C. L.</fnm></au>
    <au><snm>{Jones}</snm><fnm>C.</fnm></au>
    <au><snm>{B{\^i}rzan}</snm><fnm>L.</fnm></au>
  </aug>
  <source>\apj</source>
  <pubdate>2010</pubdate>
  <volume>720</volume>
  <fpage>1066</fpage>
  <lpage>1072</lpage>
</bibl>

<bibl id="B44">
  <title><p>{The FRI/Il Break and the Bivariate Luminosity Function in Abell
  Clusters of Galaxies}</p></title>
  <aug>
    <au><snm>{Owen}</snm><fnm>F. N.</fnm></au>
    <au><snm>{Ledlow}</snm><fnm>M. J.</fnm></au>
  </aug>
  <source>The Physics of Active Galaxies</source>
  <editor>{Bicknell}, G.~V. and {Dopita}, M.~A. and {Quinn}, P.~J.</editor>
  <series><title><p>Astronomical Society of the Pacific Conference
  Series</p></title></series>
  <pubdate>1994</pubdate>
  <volume>54</volume>
  <fpage>319</fpage>
</bibl>

<bibl id="B45">
  <title><p>{On the nature of compact steep spectrum radio
  sources.}</p></title>
  <aug>
    <au><snm>{Fanti}</snm><fnm>R.</fnm></au>
    <au><snm>{Fanti}</snm><fnm>C.</fnm></au>
    <au><snm>{Schilizzi}</snm><fnm>R. T.</fnm></au>
    <au><snm>{Spencer}</snm><fnm>R. E.</fnm></au>
    <au><cnm>{Nan Rendong}</cnm></au>
    <au><snm>{Parma}</snm><fnm>P.</fnm></au>
    <au><snm>{van Breugel}</snm><fnm>W. J. M.</fnm></au>
    <au><snm>{Venturi}</snm><fnm>T.</fnm></au>
  </aug>
  <source>\aap</source>
  <pubdate>1990</pubdate>
  <volume>231</volume>
  <fpage>333</fpage>
  <lpage>346</lpage>
</bibl>

<bibl id="B46">
  <title><p>{Making Faranoff-Riley I radio sources. I. Numerical hydrodynamic
  3D simulations of low-power jets}</p></title>
  <aug>
    <au><snm>{Massaglia}</snm><fnm>S.</fnm></au>
    <au><snm>{Bodo}</snm><fnm>G.</fnm></au>
    <au><snm>{Rossi}</snm><fnm>P.</fnm></au>
    <au><snm>{Capetti}</snm><fnm>S.</fnm></au>
    <au><snm>{Mignone}</snm><fnm>A.</fnm></au>
  </aug>
  <source>\aap</source>
  <pubdate>2016</pubdate>
  <volume>596</volume>
  <fpage>A12</fpage>
</bibl>

<bibl id="B47">
  <title><p>{Three-dimensional relativistic MHD simulations of active galactic
  nuclei jets: magnetic kink instability and Fanaroff-Riley
  dichotomy}</p></title>
  <aug>
    <au><snm>{Tchekhovskoy}</snm><fnm>A.</fnm></au>
    <au><snm>{Bromberg}</snm><fnm>O.</fnm></au>
  </aug>
  <source>\mnras</source>
  <pubdate>2016</pubdate>
  <volume>461</volume>
  <fpage>L46</fpage>
  <lpage>L50</lpage>
</bibl>

</refgrp>
} 

\appendix

\section{Dependence on the EOS}
\label{sec:eos}

\begin{figure*}
\centering
\includegraphics[width=0.6\columnwidth]{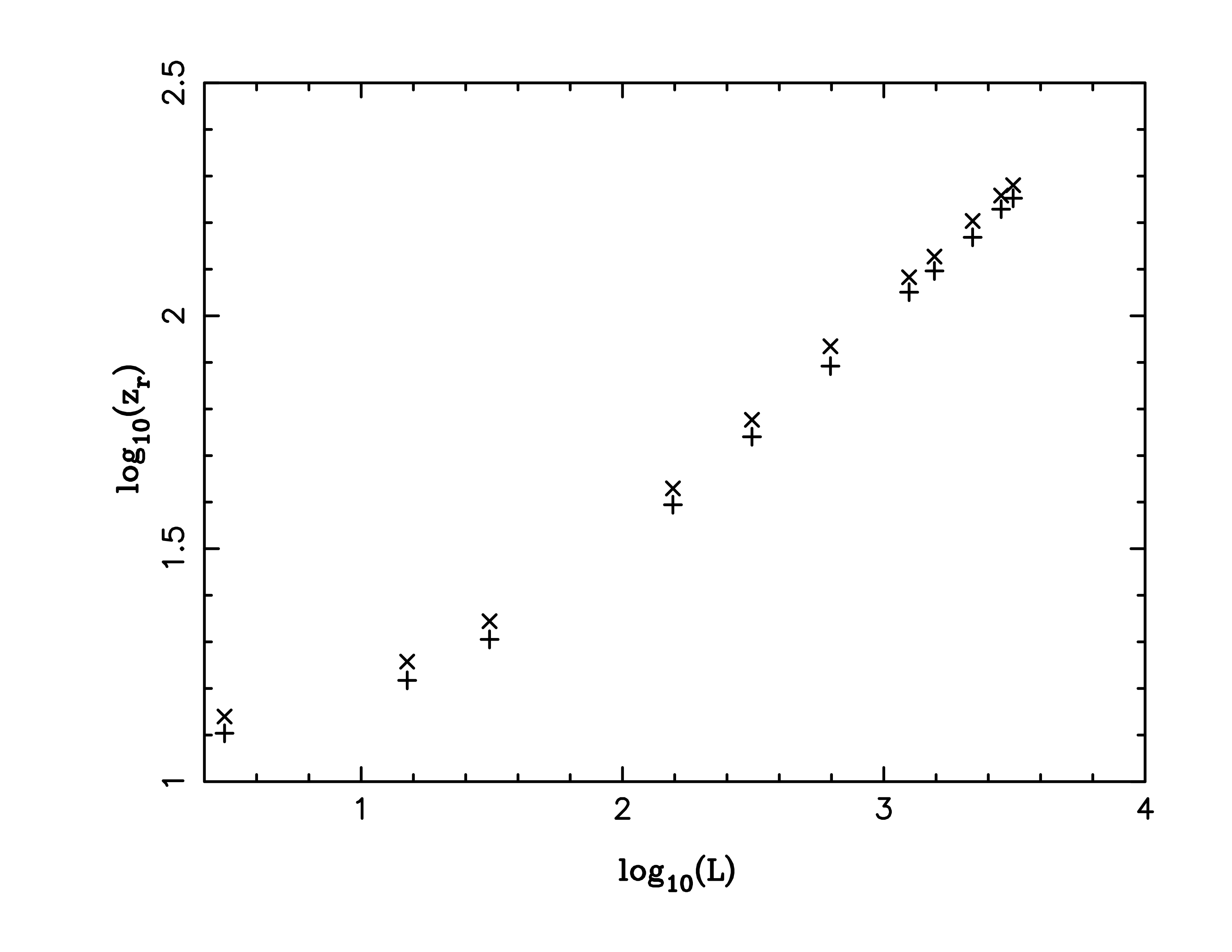}
\caption{Dependence on the ratio of specific heats. The horizontal crosses show the solutions corresponding to $\gamma=5/3$ whereas the diagonal crosses to $\gamma=4/3$. The external gas model is uniform ($\kappa=0$).   }
\label{fig:eos}
\end{figure*}

In order to probe the dependence of the jet solutions on the assumed ratio of specific heats, we compared models with  $\gamma=5/3$ and $\gamma=4/3$. Figure \ref{fig:eos} shows the results for the uniform distribution of external gas. The setup is exactly the same as in the $\kappa=0$ model discussed in Sec.\ref{sec:power-law}. One can see that the reconfinement distance is slightly longer for $\kappa=4/3$. Moreover, the difference does not vary much with the distance and remains at the level of ten percent. 

\section{Data}
\label{Data}

Tables~\ref{tabkappa0},~\ref{tabkappa1d2},~\ref{tabkappa1} and~\ref{tabkappa3d2} show the values of luminosity input into the simulations and the corresponding position of the reconfinement point  $z_\ind{r}$ for the power-law cases $\kappa=0,0.5,1$ and 1.5 respectively. We also give the reconfinement position found using the KF model, $z_\ind{r,KF}$, and the value of the quantity $A/\delta$, which expresses the effect of the inlet radius on the solution (see equation~\ref{recon}). Note that as this quantity increases the disagreement between the KF model and our computational results becomes smaller. One can check from the simulations that for increased $A/\delta$ the pressure in the post-shock gas becomes more uniform, tending to the value of the external pressure. As explained in~\cite{NS-09} the assumption of no pressure structure in the post-shock region is the main source of error in the KF model, so the stronger agreement in this limit is to be expected. Table~\ref{tabkappa2} shows the shock positions for various luminosities in the case $\kappa=2$, whilst Tables~\ref{tabzc50kappa1},~\ref{tabzc50kappa5d4},~\ref{tabzc50kappa3d2} show the King profile data in the case $z_{\ind{c}}=50$. Finally, Tables~\ref{tabzc1000kappa1} and ~\ref{tabzc1000kappa1mod} show the results for both the King profile ($\kappa=1$) and modified King profile for $z_{\ind{c}}=1000$. Clearly, since $z_{\ind{r}}<z_{\ind{c}}$ for all the luminosities chosen the results for the King case are very similar to those of the uniform profile.

 \begin{table}
\caption{Variation of axial position of the reconfinement shock with luminosity for $\kappa=0$.}
\centering
\begin{center}
\begin{tabular}{  c | c | c | c  } 
\hline
$L$ & $A/\delta$ & $z_\ind{r,KF}$ & $z_\ind{r}$\\  [0.5ex]
\cline{1-4}
 3126 & 0.119 & 94.0 & 178.8  \\ 
 2813 & 0.126 & 89.4 & 169.4 \\
 2188 & 0.142 & 79.9 & 147.4 \\ 
 1563 & 0.168 & 69.2 & 124.8 \\ 
 1250 & 0.188 & 62.9 & 112.4 \\ 
 625 & 0.266 & 47.5 & 78.0 \\ 
 313 & 0.376 & 36.5 & 55.0 \\ 
 156 & 0.533 & 28.7 & 39.3 \\ 
 31 & 1.20 & 18.4 & 20.2 \\  
 16 & 1.66 & 15.9 & 16.5 \\
  3 & 3.84 & 12.7 & 12.7 \\ 
  \hline
 \end{tabular}
\end{center}
\label{tabkappa0}
\end{table}

 \begin{table}
\caption{Variation of axial position of the reconfinement shock with luminosity for $\kappa=0.5$.}
\centering
\begin{center}
\begin{tabular}{  c | c | c | c  } 
\hline
$L$ & $A/\delta$ & $z_\ind{r,KF}$ & $z_\ind{r}$\\  [0.5ex]
\cline{1-4}
 3126 & 0.159 & 141.4 & 363.7 \\
 2813 & 0.167 & 132.7 & 338.8 \\
 2188 & 0.190 & 115.1 & 285.6 \\
 1563 & 0.225 & 95.8 & 225.7 \\
 1250 & 0.251 & 84.7 & 194.2  \\ 
 938 & 0.290 & 72.9 & 158.1 \\
 625 & 0.355 & 59.5 & 120.2 \\ 
 313 & 0.502 & 43.0 & 75.1 \\ 
 156 & 0.711 & 32.2 & 48.3 \\ 
 31 & 1.59 & 19.2 & 21.2 \\ 
 16 & 2.22 & 16.3 & 16.8 \\ 
 3 & 5.13 & 12.7 & 12.7 \\ 
 \hline 
 \end{tabular}
\end{center}
\label{tabkappa1d2}
\end{table}

\begin{table}
\caption{Variation of axial position of the reconfinement shock with luminosity for $\kappa=1$.}
\centering
\begin{center}
\begin{tabular}{  c | c | c | c  } 
\hline
$L$ & $A/\delta$ & $z_\ind{r,KF}$ & $z_\ind{r}$\\  [0.5ex]
\cline{1-4}
 1250 & 0.377 & 133.2 & 535.4  \\ 
 938 & 0.435 & 108.6 & 382.4 \\
 625 & 0.533 & 82.6 & 248.4 \\ 
 313 & 0.753 & 54.5 & 124.1 \\ 
 156 & 1.07 & 37.4 & 64.6 \\ 
 31 & 2.39 & 20.1 & 22.3 \\ 
 16 & 3.33 & 16.8 & 17.2 \\ 
 3 & 7.69 & 12.9 & 12.8 \\ 
 \hline 
 \end{tabular}
\end{center}
\label{tabkappa1}
\end{table}

 \begin{table}
\caption{Variation of axial position of the reconfinement shock with luminosity for $\kappa=1.5$.}
\centering
\begin{center}
\begin{tabular}{  c | c | c | c  } 
\hline
$L$ & $A/\delta$ & $z_\ind{r,KF}$ & $z_\ind{r}$\\  [0.5ex]
\cline{1-4}
 438 & 1.27 & 102.3 & 782.4  \\ 
 375 & 1.38 & 88.2 & 545.2 \\ 
 313 & 1.51 & 76.7 & 358.0 \\
 250 & 1.68 & 64.6 & 234.4 \\ 
 156 & 2.13 & 46.6 & 109.4 \\ 
 31 & 4.78 & 21.4 & 23.9 \\ 
 16 & 6.66 & 17.4 & 17.7 \\ 
 3 & 15.4 & 13.0 & 12.8 \\  
 \hline
 \end{tabular}
\end{center}
\label{tabkappa3d2}
\end{table}

 \begin{table}
\caption{Variation of axial position of the reconfinement shock with luminosity for $\kappa=2$.}
\centering
\begin{center}
\begin{tabular}{  c | c | c | c  } 
\hline
$L$ & $A/\delta$ & $z_\ind{r,KF}$ & $z_\ind{r}$\\  [0.5ex]
\cline{1-4} 
 125 & N/A & 53.4 & 175.1 \\
 94 & N/A & 42.6 & 84.3 \\ 
 63 & N/A & 32.7 & 46.1 \\ 
 31 & N/A & 23.1 & 25.4 \\ 
 16 & N/A & 18.0 & 18.2 \\ 
 13 & N/A & 17.0 & 16.8 \\  
 9 & N/A & 15.8 & 15.5 \\ 
 6 & N/A & 14.5 & 14.2 \\ 
 3 & N/A & 13.0 & 12.9 \\  
 \hline
 \end{tabular}
\end{center}
\label{tabkappa2}
\end{table}

 \begin{table}
\caption{Variation of axial position of the reconfinement shock with luminosity for $\kappa=1$ (King profile).}
\centering
\begin{center}
\begin{tabular}{  c | c  } 
\hline
$L$ & $z_\ind{r}$\\  [0.5ex]
\cline{1-2}
 4689 & 420.3  \\
 4376 & 389.8  \\
 3751 & 339.7  \\
 3126 & 292.4  \\
 2813 & 268.5  \\
 2188 & 216.4  \\ 
 1563 & 168.6 \\  
 938 & 119.6 \\
 625 &  92.0  \\ 
 313 & 59.5 \\ 
 156 & 40.9 \\
 63 & 26.6  \\ 
  31 & 20.3 \\  
 16 & 16.5 \\
  3 & 12.6 \\ 
  \hline
 \end{tabular}
\end{center}
\label{tabzc50kappa1}
\end{table}

 \begin{table}
\caption{Variation of axial position of the reconfinement shock with luminosity for $\kappa=1.25$ (King profile).}
\centering
\begin{center}
\begin{tabular}{  c | c  } 
\hline
$L$ & $z_\ind{r}$\\  [0.5ex]
\cline{1-2}
 4689 & 567.9  \\
 4064 & 475.0  \\
 3751 & 432.6  \\
 3438 & 393.1  \\
 3126 & 356.7  \\
 2813 & 322.7  \\
 2188 & 254.5  \\ 
 1563 & 189.8 \\  
 938 & 125.9 \\  
 313 & 60.9 \\ 
 156 & 41.4 \\
 63 & 26.7  \\ 
 31 & 20.5 \\  
 16 & 16.5 \\
  3 & 12.6 \\ 
  \hline
 \end{tabular}
\end{center}
\label{tabzc50kappa5d4}
\end{table}

\begin{table}
\caption{Variation of axial position of the reconfinement shock with luminosity for $\kappa=1.5$ (King profile).}
\centering
\begin{center}
\begin{tabular}{  c | c  } 
\hline
$L$ & $z_\ind{r}$\\  [0.5ex]
\cline{1-2}
 3438 & 559.0  \\
 3126 & 471.5  \\
 2813 & 408.8  \\
 2188 & 305.6  \\ 
1563 & 215.4 \\  
 938 & 134.6 \\
 625 & 98.5  \\  
 313 & 62.1 \\ 
 156 & 41.8 \\
 63 & 26.9  \\ 
 31 & 20.5 \\  
 16 & 16.6 \\
 3 & 12.7 \\ 
  \hline
 \end{tabular}
\end{center}
\label{tabzc50kappa3d2}
\end{table}
 
 \begin{table}
\caption{Variation of axial position of the reconfinement shock with luminosity for $\kappa=1$, $z_{\ind{c}}=1000$ (King profile).}
\centering
\begin{center}
\begin{tabular}{  c | c  } 
\hline
$L$ & $z_\ind{r}$\\  [0.5ex]
\cline{1-2}
 12503 & 363.5  \\
 9379 & 309.3  \\
 7815 & 281.6  \\
 6252 & 252.0  \\
 4689 & 218.3  \\ 
 3126 & 176.4 \\  
 1563 & 125.4 \\
 625 &  77.7  \\ 
 313 & 55.1 \\ 
 156 & 39.3 \\
  \hline
 \end{tabular}
\end{center}
\label{tabzc1000kappa1}
\end{table}

 \begin{table}
\caption{Variation of axial position of the reconfinement shock with luminosity (modified King profile).}
\centering
\begin{center}
\begin{tabular}{  c | c  } 
\hline
$L$ & $z_\ind{r}$\\  [0.5ex]
\cline{1-2}
 12503 & 361.7  \\
 9379 & 305.8  \\
 7815 & 279.3  \\
 6252 & 235.8  \\
 4689 & 186.0  \\
 3751 & 147.0  \\ 
 3126 & 124.5 \\  
 2188 & 86.5 \\
 1563 &  64.2  \\ 
 938 & 42.2 \\ 
 625 & 32.1 \\
 313 & 22.0  \\ 
 156 & 17.0 \\  
  \hline
 \end{tabular}
\end{center}
\label{tabzc1000kappa1mod}
\end{table}

\end{document}